\documentclass[11pt, letterpaper]{article}

\usepackage{amsmath, amssymb, amsfonts, amsthm}
\usepackage{mathtools} % For := and better formatting
\usepackage{bbm}       % For the indicator function \mathbbm{1}

\usepackage[margin=1in]{geometry} % Standard 1-inch margins
\usepackage{setspace}             % For line spacing (usually 1.2 or 1.5)
\usepackage{booktabs}             % For professional-looking tables
\usepackage{graphicx}             % For including plots
\usepackage[dvipsnames]{xcolor}   % For colored links

\usepackage[round]{natbib}        % "Author (Year)" style citations
\usepackage{hyperref}             % Clickable links
\hypersetup{
	colorlinks=true,
	linkcolor=NavyBlue,
	citecolor=NavyBlue,
	urlcolor=NavyBlue
}

\theoremstyle{definition}

\theoremstyle{remark}

\title{\huge Keeping Up with the Correlations \\ \LARGE Stochastic Spot/Volatility Correlation and Exotic Pricing}
\author{
	Mark Higgins \\
	\small{\texttt{mghiggins@yahoo.com}}
}
\date{\today}

\begin{document}
	
	\maketitle
	
	\begin{abstract}
        \noindent We consider a novel use case for the Double Heston model \citep{Christoffersen2009}, where the two Heston sub-variances have different spot/volatility correlations but the same volatility of volatility and mean reversion speed.
        
        This parameterization generalizes the traditional Heston stochastic volatility model \citep{Heston1993} to include stochastic spot/volatility correlation. It is an affine model, allowing European options to be priced efficiently by numerically integrating over a closed-form characteristic function.

        This model incorporates a key dynamic relevant for pricing barrier derivatives in the foreign exchange markets: a positive correlation between moves in implied volatility skew and moves in the spot price. We analyze that correlation and its impact on both barrier option pricing and volatility swap pricing. Those price impacts are comparable to or larger than the bid/ask spreads for these products. 
        
        Adding stochastic spot/volatility correlation increases the prices of out-of-the-money knockout options and one touch options, assuming that the model is calibrated to market vanilla option prices. It also increases the fair strike of volatility swaps compared to the Heston model.
	\end{abstract}

    \section{Introduction}

    The Heston stochastic volatility model is a standard extension to the Black-Scholes model \citep{BlackScholes1973} that allows the instantaneous volatility of the asset spot price to be stochastic and mean reverting, capturing an important real world dynamic for most financial markets.
		
    One Heston model parameter is the spot/volatility correlation, which defines the expected move in instantaneous volatility for a given move in the asset spot price. A positive correlation leads to an upward-sloping implied volatility skew, and vice versa.
    
    In the Heston model this correlation parameter is a constant. In practice, however, this correlation does vary over time, and moves in that correlation can themselves be well correlated with moves in the spot price.
    
    That dynamic - the correlation of the spot/volatility correlation with the spot price - is important in practice for the pricing of barrier derivatives such as knockout options and one touch options (sometimes called American digital options). This is particularly relevant in the foreign exchange options markets where that correlation is significant and barrier derivatives are relatively liquid.

    In this paper we cleanly generalize the Heston model to include stochastic spot/volatility correlation. We analyze volatility skew dynamics in the foreign exchange options markets to estimate the size of this effect. We also discuss how the pricing of barrier derivatives and volatility swaps is affected by this correlation, developing intuition for the impact based on hedging arguments and quantifying the price impact compared to the Heston model.

    An important result is that adding stochastic spot/volatility correlation increases the prices of one touch and out-of-the-money knockout options on a scale comparable to or larger than the bid/ask spread for these derivatives, assuming the model is calibrated to market vanilla option prices. Similarly, it meaningfully increases the fair strikes of volatility swaps. Market makers using the Heston model to price and manage risk for these derivatives may find themselves selling these derivatives at below-market prices and accumulating mispriced short positions.
	
	\section{The Model}
	
    We define a generalization of the Heston stochastic volatility model as follows:
	
	\begin{align*}
		\frac{dS_t}{S_t} &= (r - q) \,dt + \sqrt{v_t} \,dW^S_t \\
        \sqrt{v_t} dW^S_t &= \sqrt{v^+_t} dW^{S+}_t + \sqrt{v^-_t} dW^{S-}_t \\
        v_t &= v^+_t + v^-_t \\
		dv^+_t &= \beta (\theta_+ - v^+_t) \,dt + \alpha \sqrt{v^+_t} \,dW^+_t \\
		dv^-_t &= \beta (\theta_- - v^-_t) \,dt + \alpha \sqrt{v^-_t} \,dW^-_t \\
        d\langle W^{S+}, W^+ \rangle_t &= (\bar\rho + \eta) \,dt \\
        d\langle W^{S-}, W^- \rangle_t &= (\bar\rho - \eta) \,dt \\
        d\langle W^+, W^- \rangle_t &= d\langle W^{S+}, W^{S-}\rangle_t = d\langle W^{S+}, W^- \rangle_t = d\langle W^{S-}, W^+ \rangle_t = 0
	\end{align*}

	$S_t$ is the asset spot price (``spot'') and $v_t$ is the instantaneous volatility squared, often called the ``variance'', which equals the sum of the two sub-variance processes $v^+_t$ and $v^-_t$. Those two processes share a mean reversion strength $\beta$ and volatility of volatility $\alpha$, but have different spot/volatility correlations. $v^+$ has a more positive correlation, equal to the average correlation $\bar\rho$ plus the correlation half-range $\eta$. $v^-$ has a more negative correlation, equal to the average correlation $\bar\rho$ minus $\eta$. Naturally, $\bar\rho - \eta>-1$ and $\bar\rho + \eta < 1$ to keep all correlations in $(-1,1)$.

    This is how correlation becomes stochastic in this model: by changing the weights of the two sub-variance processes and thereby letting the correlation range between $\bar\rho - \eta$ and $\bar\rho + \eta$.

    This model naturally induces a positive correlation between moves in spot and moves in the spot/volatility correlation: as spot increases, the sub-variance process with the higher correlation will tend to increase more than the sub-variance process with the lower correlation, and the instantaneous spot/volatility correlation will get closer to the higher correlation.
	
    As the parameter $\eta$ goes to zero the model reduces to the original Heston stochastic volatility model with constant spot/volatility correlation $\bar\rho$. In this sense, $\eta$ is a kind of volatility of correlation.

    We can also rewrite the SDE for $\frac{dS_t}{S_t}$ in terms of just three uncorrelated Brownian motions $W^0_t$, $W^+_t$, and $W^-_t$:

    \begin{align*}
    \frac{dS_t}{S_t} &= (r - q) \,dt \\
    &+ \left( (\bar\rho+\eta)\sqrt{v_t^+}\,\mathrm dW_t^{+}
    +(\bar\rho-\eta)\sqrt{v_t^-}\,\mathrm dW_t^{-}
    +\sqrt{(1-(\bar\rho+\eta)^2)v_t^+ + (1-(\bar\rho-\eta)^2)v_t^-}\,\mathrm dW_t^{0} \right)
    \end{align*}

    which is convenient for Monte Carlo simulation and demonstrates the three driving factors in this model.

    Note that this is a version of the Double Heston model \citep{Christoffersen2009}. That model was designed for matching a full term structure of implied volatilities because the two sub-variances have different mean reversion speeds and affect different parts of the term structure. In this case we have the same mean reversion speed for both sub-variances, but different spot/volatility correlations.

    \section{The Variance Process}

    Our model defines the processes for the two sub-variances; we can use those to write down the SDE for the total variance $v_t=v_t^++v_t^-$:

    \begin{align*}
    \mathrm dv_t &= dv^+_t + dv^-_t \\
    &= \beta (\theta_+ - v_t^+) \,dt + \alpha \sqrt{v^+_t} \,dW^+_t + \beta (\theta_- - v_t^-) \,dt + \alpha \sqrt{v^-_t} \,dW^-_t \\
    &= \beta (\theta - v_t) \,dt + \alpha \left(\sqrt{v^+_t} \,dW^+_t + \sqrt{v^-_t} \,dW^-_t\right) \\
    &= \beta (\theta - v_t) \,dt + \alpha \sqrt{v_t} \,dW_t^{v}
    \end{align*}

    where $\theta = \theta_+ + \theta_-$ is the long run variance level. This is exactly the Heston/CIR variance SDE, so this model reproduces the Heston model's marginal variance dynamics.
    
    \section{The Spot/Volatility Correlation}

    A key feature of our model is the stochastic spot/volatility correlation. To calculate that correlation, we first compute the instantaneous covariation between $X_t=\ln S_t$ and the total variance $v_t=v_t^++v_t^-$:

    $$
    \begin{aligned}
    \mathrm d\langle X,v\rangle_t
    &= \mathrm d\langle X, v^+ + v^-\rangle_t
    = \mathrm d\langle X,v^+\rangle_t + \mathrm d\langle X,v^-\rangle_t\\
    &= \alpha(\bar\rho+\eta)v_t^+\,\mathrm dt + \alpha(\bar\rho-\eta)v_t^-\,\mathrm dt.
    \end{aligned}
    $$

    Define the spot/variance covariance factor
    $$
    c_t := (\bar\rho+\eta)v_t^+ + (\bar\rho-\eta)v_t^- = \bar\rho\,v_t + \eta (v_t^+-v_t^-)
    = \bar\rho\,v_t + \eta u_t,
    $$
    so that
    $$
    \mathrm d\langle X,v\rangle_t = \alpha c_t\,\mathrm dt.
    $$

    Define the instantaneous spot/volatility correlation (between $W^S$ and the Brownian driver of $v_t$) by the identity
    
    $$
    \mathrm d\langle X,v\rangle_t = \alpha v_t \rho_t\,\mathrm dt,
    $$
    which gives
    \begin{equation}
    \label{eq:rho_t}
    \boxed{
    \rho_t = \frac{c_t}{v_t}
    = \bar\rho + \eta\,\frac{u_t}{v_t}
    =\bar\rho + \eta\,\frac{v_t^+ - v_t^-}{v_t^+ + v_t^-}.}
    \end{equation}

    That is, when the difference between the two sub-variance processes is zero, $\rho_t = \bar\rho$. When $v^+ \gg v^-$, $\rho_t = \bar\rho + \eta$, and when $v^- \gg v^+$, $\rho_t = \bar\rho - \eta$. So the spot/volatility correlation is constrained to the range $\bar\rho - \eta$ to $\bar\rho + \eta$, as the sub-variances are always non-negative.
    
    \section{Model Parameterization and Correlation Structure}

    There are two ways in this model to define the correlation structure: $\bar\rho$ and $\eta$, which control the allowed range of the spot/volatility correlation; and $\theta_+$ and $\theta_-$, which control the mean reversion levels of the two sub-variance processes. The long run values of $v^+$ and $v^-$ are $\theta_+$ and $\theta_-$, respectively, so the long run correlation is

    \[
    \rho_a = \bar\rho + \eta \frac{(\theta_+ - \theta_-)}{\theta}
    \]

    The instantaneous correlation is

    \[
    \rho_0 = \bar\rho + \eta \frac{(v^+_0 - v^-_0)}{v_0}
    \]

    where $v^+_0$ and $v^-_0$ are the initial values of the two sub-variance processes. On expectation $\rho_t$ varies between these two levels, starting at $\rho_0$ and mean reverting to $\rho_a$ (not $\bar\rho$).

    In this paper we look at calibrating just to a single expiration tenor at a time, so we are less concerned with the term structure of correlation. For this purpose we set $v^+_0 = \theta_+$ and $v^-_0 = \theta_-$ so that the initial correlation matches the long run correlation.

    We choose then to parameterize the model in terms of $\theta=\theta_+ + \theta_-$, $\rho_a$, and $\rho_0$, from which we can calculate $\theta_+$, $\theta_-$, $v^+_0$, and $v^-_0$. This is a more natural parameterization: $\sqrt{\theta}$ is the long run volatility level, $\rho_a$ is the long run correlation level, and $\rho_0$ is the initial correlation level. We generally choose $\bar\rho = \rho_a$, and when calibrating to a single expiration tenor assume $\rho_0 = \rho_a$ also, which implies $\theta_+=\theta_-=v^+_0=v^-_0=\theta/2$. This parameterization centers the allowed correlation range at the initial correlation level and ensures that in the $\eta \rightarrow 0$ limit the current correlation is always in the allowed range.

    \section{\texorpdfstring{The Stochastic Differential Equation for $\rho_t$}{The SDE for rho t}}
    
    We can apply Ito's Lemma to equation \ref{eq:rho_t} for $\rho_t$ to find the SDE that $\rho_t$ satisfies. Write $v_t=v_t^+ + v_t^-$ and $u_t=v_t^+ - v_t^-$. Then
    \[
    \rho_t=\bar\rho+\eta\,\frac{u_t}{v_t}
    \qquad\Longrightarrow\qquad
    \frac{u_t}{v_t}=\frac{\rho_t-\bar\rho}{\eta}.
    \]
    Using the (independent) Brownian drivers $(W^+,W^-)$ for $(v^+,v^-)$, one finds
    \[
    \mathrm dv_t=\beta(\theta-v_t)\,\mathrm dt+\alpha\left(\sqrt{v_t^+}\,\mathrm dW_t^+ + \sqrt{v_t^-}\,\mathrm dW_t^-\right),
    \quad
    \mathrm du_t=\beta\left((\theta_+-\theta_-)-u_t\right)\,\mathrm dt+\alpha\left(\sqrt{v_t^+}\,\mathrm dW_t^+ - \sqrt{v_t^-}\,\mathrm dW_t^-\right)
    \]
    Applying Ito to $u_t/v_t$ gives a drift
    \[
    \frac{\beta\theta}{v_t}(\rho_a-\rho_t)\,\mathrm dt
    \]
    where we used $\theta_+-\theta_-=\theta\,(\rho_a-\bar\rho)/\eta$, and an instantaneous variance
    \[
    \mathrm d\langle \rho,\rho\rangle_t = \frac{\alpha^2}{v_t}\left(\eta^2-(\rho_t-\bar\rho)^2\right)\,\mathrm dt.
    \]
    Therefore, there exists a Brownian motion $W^{\rho}$ such that
    \begin{equation}
      \label{eq:rho_sde}
    \boxed{
        \mathrm d\rho_t
        =
        \beta \frac{\theta}{v_t} (\rho_a - \rho_t) \,\mathrm dt
        +\frac{\alpha}{\sqrt{v_t}}\sqrt{\eta^2-(\rho_t-\bar\rho)^2}\,\mathrm dW_t^{\rho}
    }
	\end{equation}
	
    This looks like a process that mean reverts to $\rho_a$ with a variable speed determined by $\beta \frac{\theta}{v_t}$; thus, the mean reversion speed for $\rho_t$ is quite similar to the volatility mean reversion speed. When $\rho_t=\bar\rho$, the "volatility of correlation" is $\frac{\alpha \eta}{\sqrt{v_t}}$ - that is, proportional to $\eta$.

    This process constrains $\rho_t$ to lie in the range $\bar\rho - \eta$ to $\bar\rho + \eta$ - its volatility goes to zero at those edges - but that was already clear from the definition of $\rho_t$.

	\section{Spot/Rho Correlation}
	
	The key market dynamic that we are introducing with this model is the correlation of moves in the asset spot price with moves in the spot/volatility correlation. This ``spot/rho correlation'' $\rho_{cs}$ is positive in this model: as spot increases, the sub-variance process with the more positive correlation tends to increase in comparison to the one with more negative correlation, and the instantaneous correlation (per equation \ref{eq:rho_t}) tends to turn more positive as well.
	
	We can use the SDE for $\rho_t$ from equation \ref{eq:rho_sde} to quantify the covariance between moves in the log spot price $X_t = \ln(S_t)$ and moves in $\rho_t$:
	
	\[
	d\langle X, \rho \rangle_t = \alpha (\eta^2 - (\bar\rho - \rho_t)^2) dt
    \]
	
	And the instantaneous variance of $\rho_t$ is
	
	\[
	d\langle \rho, \rho \rangle_t = \frac{\alpha^2}{v_t} (\eta^2 - (\bar\rho - \rho_t)^2) dt
	\]
	
	The instantaneous variance of $X_t$ is just $v_t dt$. Therefore,
	
	\begin{equation}
    \label{eq:rho_cs}
	\boxed{
		\rho_{cs} = \mathrm{Corr}(\mathrm dX_t,\mathrm d\rho_t) = \sqrt{\eta^2 - (\bar\rho - \rho_t)^2}
	}
	\end{equation}
	
	When $\rho_t=\bar\rho$ this simplifies to $\eta$: this parameter controls both the volatility of correlation and the spot/rho correlation.
	
    \section{European Vanilla Option Pricing}
    \label{sec:vanilla_pricing}
	
	We will follow the same approach to pricing European vanilla prices as in the Heston model derivation, by finding a closed-form expression for the characteristic function of $X_T=\ln(S_T)$ on the option expiration date $T$.
	
    \subsection{\texorpdfstring{Affine Generator in State $(x,v^+,v^-)$}{Affine Generator in State (x, v+, v-)}}

    Let $X_t=\ln S_t$. Using the independent Brownian drivers $(W^+,W^-,W^0)$ and the dynamics above, the nonzero instantaneous covariations involving the state are:
    $$
    \mathrm d\langle X,X\rangle_t = (v^+ + v^-)\mathrm dt,\quad
    \mathrm d\langle v^+,v^+\rangle_t = \alpha^2 v^+ \mathrm dt,\quad
    \mathrm d\langle v^-,v^-\rangle_t = \alpha^2 v^- \mathrm dt,
    $$
    $$
    \mathrm d\langle X,v^+\rangle_t = \alpha(\bar\rho+\eta)v^+ \mathrm dt,\quad
    \mathrm d\langle X,v^-\rangle_t = \alpha(\bar\rho-\eta)v^- \mathrm dt.
    $$
    
    Hence the backward operator $\mathcal L$ acting on smooth $f(x,v^+,v^-)$ is
    $$
    \begin{aligned}
    \mathcal L f
    &= \left(r-q-\tfrac12(v^+ + v^-)\right)f_x
    + \beta(\theta_+ - v^+)f_{v^+} + \beta(\theta_- - v^-)f_{v^-}\\
    &\quad + \tfrac12(v^+ + v^-) f_{xx}
    + \tfrac12\alpha^2 v^+ f_{v^+v^+} + \tfrac12\alpha^2 v^- f_{v^-v^-}\\
    &\quad + \alpha(\bar\rho+\eta)v^+ f_{x v^+} + \alpha(\bar\rho-\eta)v^- f_{x v^-}.
    \end{aligned}
    $$
    All coefficients are affine in $(v^+,v^-)$.
    
    \subsection{Exponential-Affine Ansatz and Riccati System}
    
    For $\xi\in\mathbb C$, define the characteristic function as
    $$
    \Phi(t,x,v^+,v^-;\xi)
    =\mathbb E\!\left[e^{i \xi X_T}\mid X_t=x,\ v_t^+=v^+,\ v_t^-=v^-\right]
    $$

    and define the remaining time to expiration $\tau=T-t$.
    Use the affine form
    $$
    \Phi = \exp\!\left(A(\tau;\xi)+B_+(\tau;\xi)v^+ + B_-(\tau;\xi)v^- + i \xi x\right),
    \qquad A(\tau=0)=B_+(0)=B_-(0)=0.
    $$
    
    Matching coefficients yields the Riccati ODEs:
    $$
    \begin{aligned}
    A'(\tau)
    &= i \xi (r-q) + \beta\theta_+\,B_+(\tau) + \beta\theta_-\,B_-(\tau),\\[3pt]
    B_+'(\tau)
    &= -\tfrac12(\xi^2+i \xi) - \beta B_+(\tau)
    + \tfrac12\alpha^2 B_+(\tau)^2
    + i \xi \,\alpha(\bar\rho+\eta)\,B_+(\tau),\\[3pt]
    B_-'(\tau)
    &= -\tfrac12(\xi^2+i \xi) - \beta B_-(\tau)
    + \tfrac12\alpha^2 B_-(\tau)^2
    + i \xi \,\alpha(\bar\rho-\eta)\,B_-(\tau).
    \end{aligned}
    $$

    \subsection{Closed-Form Solution to the Riccati ODEs}
    
    This system of ODEs can be solved analytically to give the characteristic function. Appendix \ref{app:ode} contains the full derivation, but the results are given below.

    Define the two limiting correlations

    $$
    \rho_+ := \bar\rho+\eta,\qquad \rho_- := \bar\rho-\eta.
    $$

    and define the following complex constants, depending on $\xi$, the Fourier argument of the characteristic function:

    $$
    b_\pm(\xi) := \beta - i\xi\,\alpha\,\rho_\pm,
    \qquad
    d_\pm(\xi) := \sqrt{b_\pm(\xi)^2+\alpha^2(\xi^2+i\xi)},
    \qquad
    g_\pm(\xi) := \frac{b_\pm(\xi)-d_\pm(\xi)}{b_\pm(\xi)+d_\pm(\xi)}.
    $$

    then the characteristic function is given by

    \begin{equation}
    \boxed{%
        \begin{aligned}
        A(\tau;\xi)
        &= i\xi(r-q)\tau
        +\frac{\beta}{\alpha^2 }\sum_{\pm}
        \theta_\pm \left[
        (b_\pm(\xi)-d_\pm(\xi))\,\tau
        -2 \ln\!\left(\frac{1-g_\pm(\xi)e^{-d_\pm(\xi)\tau}}{1-g_\pm(\xi)}\right)
        \right] \\
        B_\pm(\tau;\xi)
        &= \frac{b_\pm(\xi)-d_\pm(\xi)}{\alpha^2}\,
        \frac{(1-e^{-d_\pm(\xi)\tau})}{(1-g_\pm(\xi)e^{-d_\pm(\xi)\tau})},
        \end{aligned}%
    }
    \end{equation}

    and

    \begin{equation}
    \boxed{
        \Phi(t,x,v^+,v^-;u) = \mathbb E\!\left[e^{iu\ln S_T}\mid\mathcal F_t\right]
    =
    \exp\!\left(A(\tau;u)+B_+(\tau;u)v_t^+ + B_-(\tau;u)v_t^- + iu\ln S_t\right).
    }
    \end{equation}

    Given the characteristic function, European vanilla option prices can be computed efficiently via Fourier inversion; a standard approach is the Carr-Madan FFT method \citep{CarrMadan1999}.

    \section{Model Implied Volatilities}

    Adding stochastic spot/volatility correlation tends to increase the implied volatility smile beyond the smile that comes from pure stochastic volatility. That happens because we expect the skew to get more positive as spot goes up, so instantaneous volatility rises relatively faster than a pure Heston model as spot increases (and vice versa as spot decreases).

\begin{figure}[htbp]
		\centering
		\includegraphics[width=0.7\textwidth]{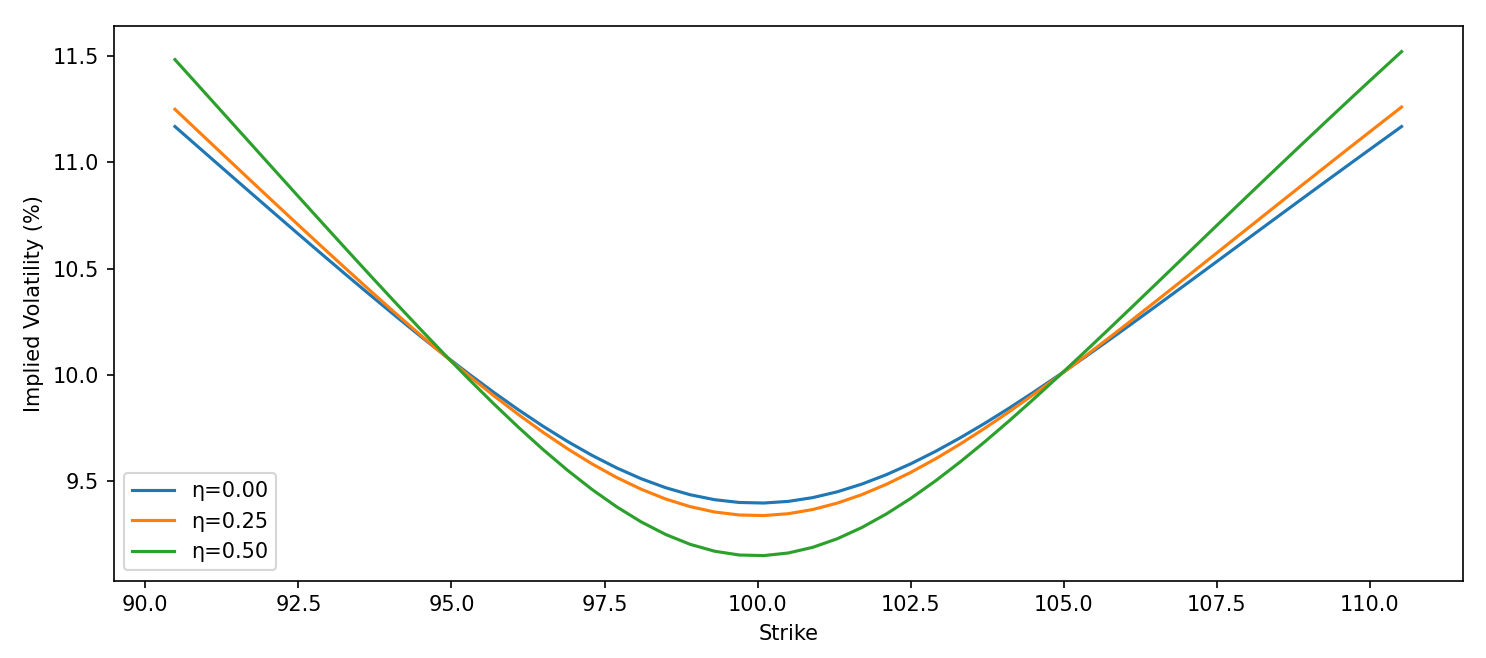}
		\caption{Implied Volatility Impact of Adding Stochastic Spot/Volatility Correlation}
		\label{fig:imp_vols}
		\vspace{0.5em}
		\begin{minipage}{0.9\textwidth}
			\footnotesize
			Increasing $\eta$ tends to increase the implied volatility smile beyond the smile that comes from pure stochastic volatility.  The $\eta=0$ limit is the Heston model.
		\end{minipage}
	\end{figure}

    Figure \ref{fig:imp_vols} shows a chart of the implied volatility smile for a simplified set of parameters: $S=100$, $r=q=0$, $T=0.25$, $\theta=0.01$, $\alpha=0.3$, $\beta=2$, and $\bar\rho=\rho_a=\rho_0=0$. As $\eta$ increases, the smile gets more pronounced. Note that the $\eta=0$ limit is the Heston model, with constant spot/volatility correlation.

    \section{Model Calibration}
	
	One important use of this model is pricing exotic derivatives like knockout options. For that purpose it is important that the model reliably reproduce the prices of benchmark vanilla options traded in the market so that the model is accurately representing the cost of hedges in the market.
	
	We consider calibration to three implied volatilities on a single expiration date, and, following FX market conventions, choose to calibrate to the at-the-money implied volatility, the 25-delta risk reversal, and the 25-delta butterfly. The risk reversal is a measure of volatility skew: the 25-delta risk reversal is the implied volatility of a high strike call option whose delta equals 0.25, less the implied volatility of a low strike put option whose delta equals -0.25. The butterfly is a measure of smile, and the 25-delta butterfly equals the average of those two out-of-the-money option implied volatilities, less the at-the-money volatility. Given the at-the-money volatility, the 25-delta risk reversal, and the 25-delta butterfly, one can invert to calculate the implied volatilities of the high- and low-strike options.
	
	As we have three input implied volatilities to calibrate the model to, we must choose just three model parameters to participate in the calibration. For the Heston stochastic volatility model, $v_0$ and $\theta$ mostly determine the at-the-money volatility - generally these are set to the same value when calibrating to a single expiration, since their difference mostly affects the term structure of at-the-money implied volatility. The value of $\alpha$ mostly determines the level of the butterfly, and the product of $\alpha$ and $\rho$ determine the level of the risk reversal. Typically $\beta$ is specified a priori for single-expiration calibrations: it affects the term structure of at-the-money volatility as well as the butterfly. We are doing single-expiration calibrations so we do not care about the term structure of at-the-money volatility, and we can use $\alpha$ to match the level of the butterfly.
	
	We will use a similar approach for calibrating our model parameters. The parameters we calibrate are $\theta(=v_0)$, $\alpha$, and $\bar\rho(=\rho_a=\rho_0)$. The parameters we fix a priori are $\beta$ and $\eta$.
	
	We consider an example market for options with 0.25y to expiration, at-the-money volatility 8.0\%, 25-delta risk reversal +1.0\%, and 25-delta butterfly +0.5\%. We use $\beta=2$ as a representative value. We then calibrate the three parameters such that the model implied volatilities match the market implied volatilities at the three strikes (25-delta put strike, at-the-money strike, and 25-delta call strike).
	
\begin{figure}[htbp]
		\centering
		\includegraphics[width=0.7\textwidth]{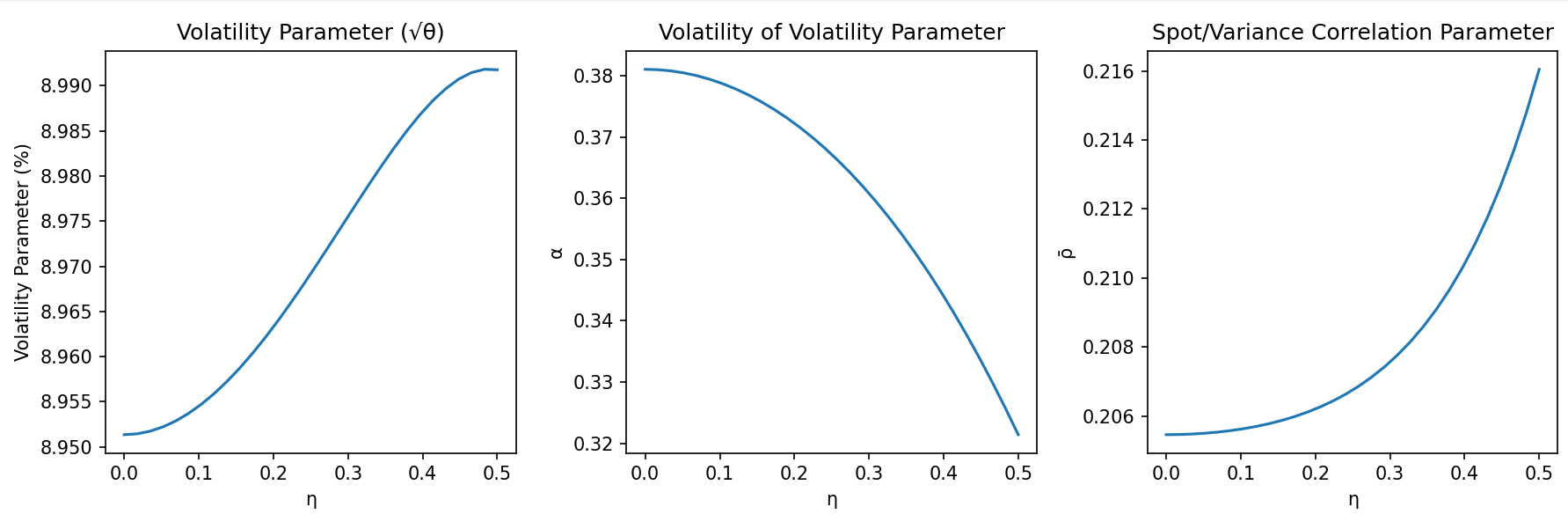}
		\caption{Calibrated Parameters vs Volatility of Correlation}
		\label{fig:calibparams}
		\vspace{0.5em}
		\begin{minipage}{0.9\textwidth}
			\footnotesize
			Increasing $\eta$ has noticeable but relatively limited impact on the calibrated values of $\theta$ and $\bar\rho=\rho_a$. However, it has a larger impact on the volatility of volatility parameter $\alpha$, which decreases as $\eta$ increases.
		\end{minipage}
	\end{figure}

	Figure \ref{fig:calibparams} shows a chart of the values of the three calibrated parameters, recalibrating at different values of $\eta$ so that the model implied volatilities continue to match the market implied volatilities.
	
	Increasing $\eta$ has only a small impact on the overall volatility level, $\theta$. It has a larger effect on the calibrated $\bar\rho$ correlation level, but still relatively modest.
	
	However, it has a much larger effect on the calibrated $\alpha$ values: as we saw above, the stochastic correlation contributes to the smile, so less volatility of volatility is needed to match a given implied volatility smile. This is an interesting effect, since there are well known issues with Heston-style models when the $\alpha$ values get so big that $v=0$ becomes accessible \citep{LordKoekkoekVanDijk2010} - violating the Feller condition $\frac{\alpha^2}{2 \beta \theta}<1$. Adding stochastic correlation means that the model parameters are less likely to break the Feller threshold, making the distribution of $v_t$ more stable.
	
	\section{The Risk Reversal Beta and Parameter Estimation}
	
    \subsection{The Risk Reversal Beta Empirically}

	An important market dynamic that this model incorporates is the correlation between moves in the spot/volatility correlation and spot returns.
	
	The spot/volatility correlation is not directly observable in the market. A reasonable proxy, however, is the implied volatility skew, quantified as the risk reversal\footnote{If a delta is not specified for a risk reversal we assume it is a 25-delta risk reversal.} in foreign exchange markets. There is a separate risk reversal for each option expiration tenor quoted in the market.
	
	In the Heston stochastic volatility model, as in most stochastic volatility models, the risk reversal is well approximated as proportional to the spot/volatility correlation. That suggests that we can get an estimate of the spot/volatility correlation dynamics by looking at the dynamics of market risk reversal levels.

\begin{figure}[htbp]
		\centering
		\includegraphics[width=0.7\textwidth]{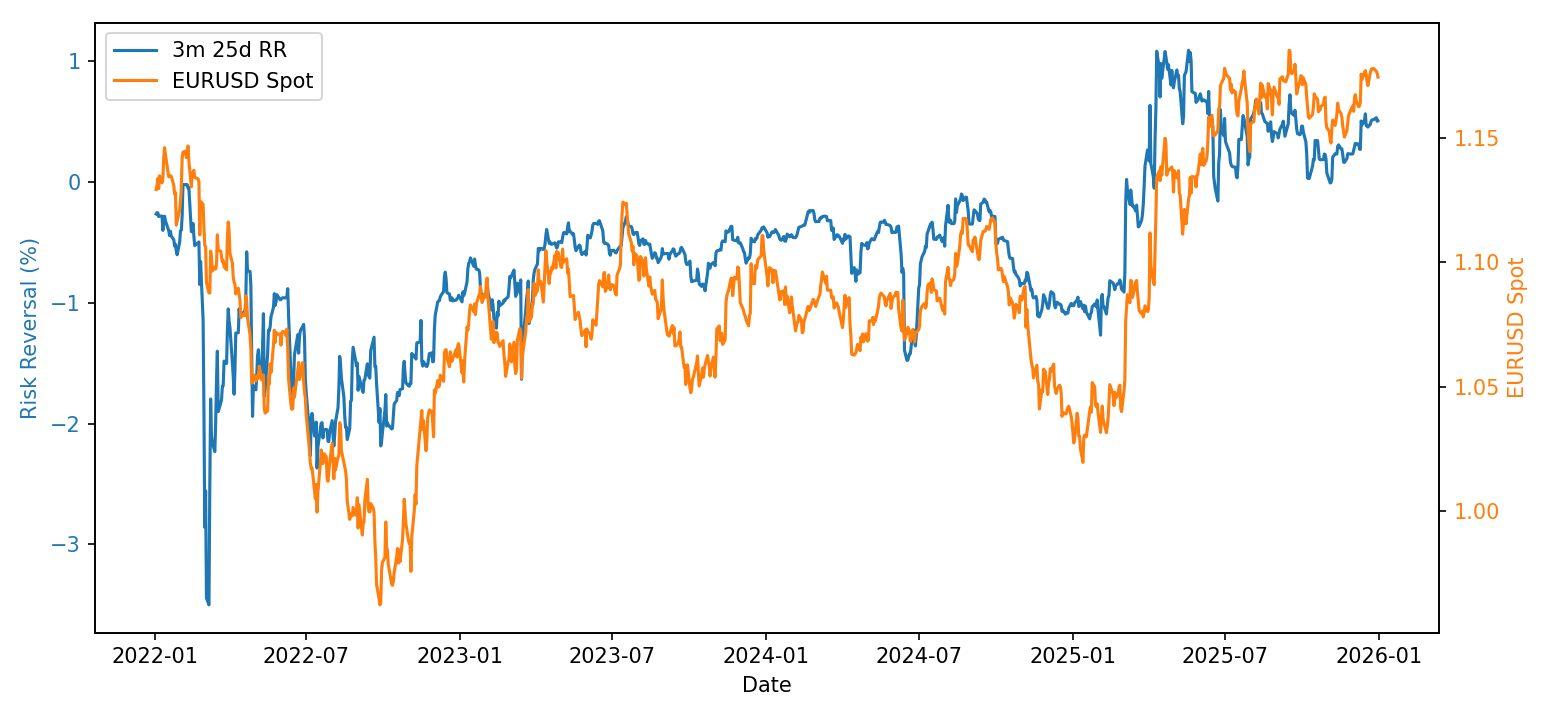}
		\caption{EURUSD 3m-Expiration 25-Delta Risk Reversals and EURUSD Spot}
		\label{fig:rrs_eurusd}
		\vspace{0.5em}
		\begin{minipage}{0.9\textwidth}
			\footnotesize
			EURUSD 3m 25-delta risk reversal (left) and EURUSD spot (right) between 2022 and 2025. Note the strong relationship between the two.
		\end{minipage}
	\end{figure}

\begin{figure}[htbp]
		\centering
		\includegraphics[width=0.7\textwidth]{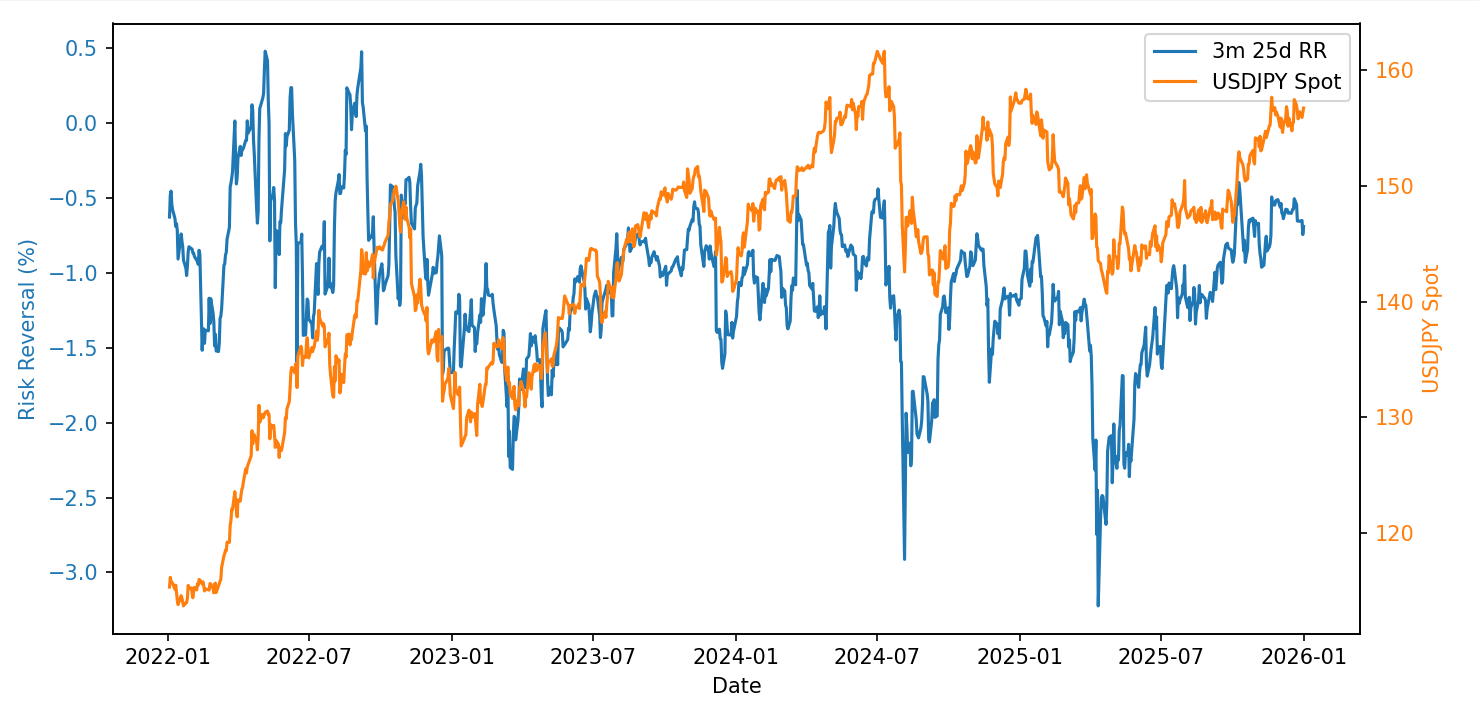}
		\caption{USDJPY 3m-Expiration 25-Delta Risk Reversals and USDJPY Spot}
		\label{fig:rrs_usdjpy}
		\vspace{0.5em}
		\begin{minipage}{0.9\textwidth}
			\footnotesize
			USDJPY 3m 25-delta risk reversal (left) and USDJPY spot (right) between 2022 and 2025. The relationship is not as consistent over time as it is for EURUSD but day to day changes are still quite correlated.
		\end{minipage}
	\end{figure}

	Figure \ref{fig:rrs_eurusd} shows a time series of EURUSD 3m-expiration risk reversal on the left axis, and EURUSD spot on the right axis, between 2022 and 2025\footnote{Data courtesy of JPMorgan Chase's foreign exchange options trading desk: London 11am representative daily closes}. Figure \ref{fig:rrs_usdjpy} shows the same for the USDJPY currency pair.
	
	\begin{figure}[htbp]
		\centering
		\includegraphics[width=0.7\textwidth]{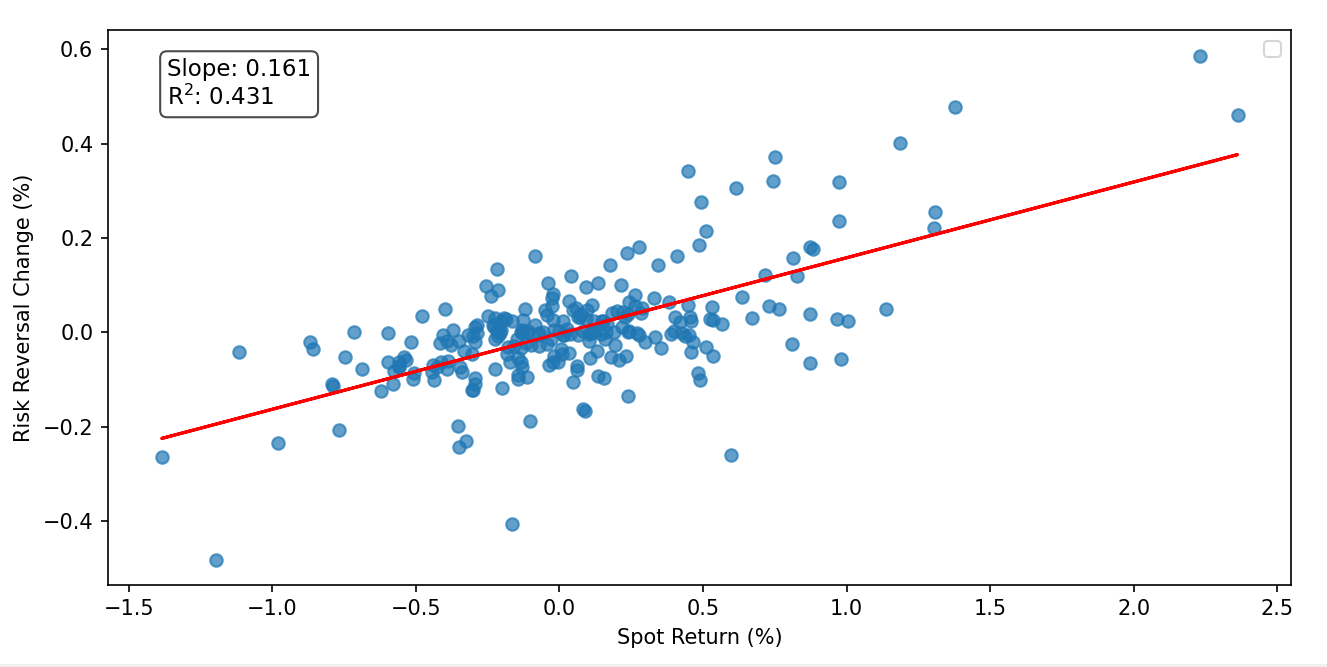}
		\caption{EURUSD 3m-Expiration Risk Reversal Beta}
		\label{fig:rrbeta_eurusd}
		\vspace{0.5em}
		\begin{minipage}{0.9\textwidth}
			\footnotesize
			A regression of daily changes in EURUSD 3m 25-delta risk reversals vs daily spot log returns using daily data from 2025. The $R^2$ of the fit is 43\%, giving a spot/risk reversal correlation of 66\%. The risk reversal beta is the slope of the line: the risk reversal tends to get more positive by 0.16\% for every 1\% increase in spot.
		\end{minipage}
	\end{figure}
	
	\begin{figure}[htbp]
		\centering
		\includegraphics[width=0.7\textwidth]{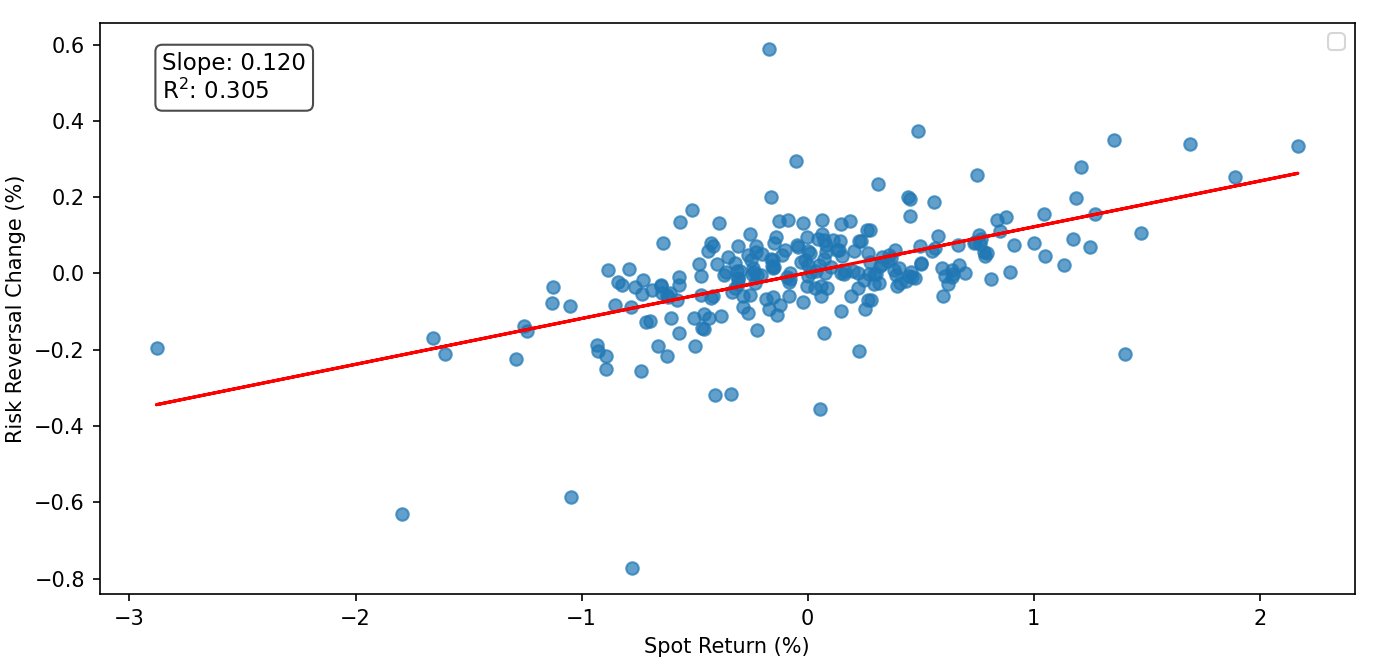}
		\caption{USDJPY 3m-Expiration Risk Reversal Beta}
		\label{fig:rrbeta_usdjpy}
		\vspace{0.5em}
		\begin{minipage}{0.9\textwidth}
			\footnotesize
			A regression of daily changes in USDJPY 3m 25-delta risk reversals vs daily spot log returns using daily data from 2025. The $R^2$ of the fit is 31\%, giving a spot/risk reversal correlation of 55\%. The risk reversal beta is the slope of the line: the risk reversal tends to get more positive by 0.12\% for every 1\% increase in spot.
		\end{minipage}
	\end{figure}
	
	Figures \ref{fig:rrbeta_eurusd} and \ref{fig:rrbeta_usdjpy} show linear regressions of daily change in 3m 25-delta risk reversal vs daily spot log returns. These figures highlight the strong dependence between these two market factors: the $R^2$ for the EURUSD regression is 43\%, translating to a spot/rho correlation of 66\%; for USDJPY the $R^2$ is 31\%, leading to a spot/rho correlation of 55\%. These regressions use a one year window, including all daily market data in 2025.
		
	\begin{figure}[htbp]
	\centering
	\includegraphics[width=0.7\textwidth]{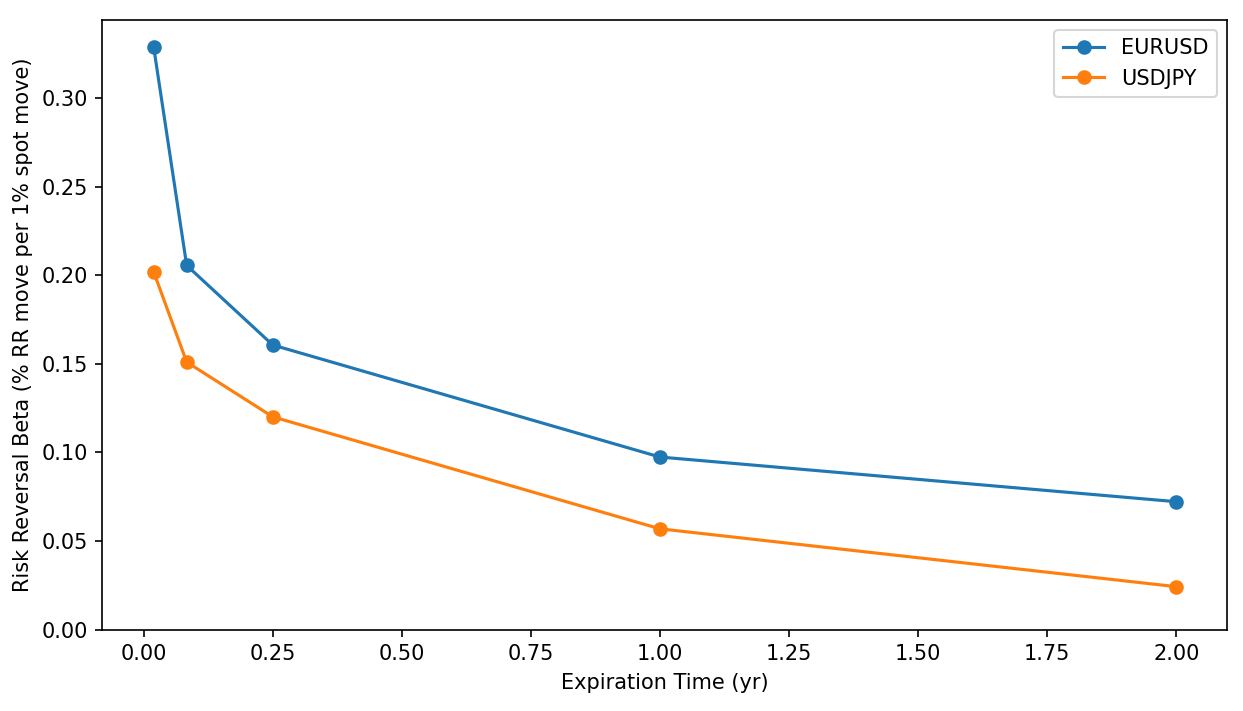}
	\caption{Empirical EURUSD and USDJPY Risk Reversal Beta by Expiration Tenor}
	\label{fig:rrbeta_by_tenor}
	\vspace{0.5em}
	\begin{minipage}{0.9\textwidth}
		\footnotesize
		Risk reversal betas tend to be larger for shorter expirations, which suggests mean reversion in the underlying spot/volatility correlation process. These values were calculated as the slopes of linear regressions of daily changes in 25-delta risk reversals vs daily spot log returns, using daily data from 2025.
	\end{minipage}
    \end{figure}
	
	The slopes of those regression lines are exactly what we define as the ``risk reversal beta''\footnote{``Risk reversal beta'' is not a standard term in the foreign exchange markets.}: how much we expect the risk reversal to move given a small move in spot. For EURUSD that risk reversal beta is 0.16 (for 3m-expiration options, regressed over data in the 2025 calendar year) - that is, the EURUSD 3m 25-delta risk reversal is expected to increase 0.16\% for every 1\% increase in spot. Note that these are absolute, not relative, changes in risk reversals, whose values happen to be quoted in percent volatilities. For USDJPY the risk reversal beta was estimated as 0.12 in this dataset.
	
	We are using 3m-expiration data as an example, but there is a different risk reversal beta for each expiration. Figure \ref{fig:rrbeta_by_tenor} shows a chart of the EURUSD and USDJPY risk reversal betas as a function of expiration tenor, calculated from linear regressions of 25-delta risk reversal changes vs spot returns (using the same daily data from 2025).\footnote{The risk reversal beta is also a function of delta. The 10-delta risk reversal beta is almost always the 25-delta risk reversal beta multiplied by a scale factor greater than 1.} Note that risk reversal beta declines with tenor, which is what we would expect from a mean reverting spot/volatility correlation process.

    \subsection{Model Risk Reversal Beta}
	
	Now we turn to calculating our model's risk reversal beta, which will help us estimate $\eta$. In this model the expression for the risk reversal beta is a somewhat complicated function of the model parameters and FX options market conventions for how to calculate a strike from a benchmark delta. However, in practice the model risk reversal value's dependence on $\rho_0$ - the initial spot/volatility correlation - is very close to linear. The slope of that line depends on the model parameters and the option expiration.
	
	We write
	
	\[
	dRR(\tau) = k(\tau) d\rho_t
	\]
	
	to express this dependence, where $k(\tau)$ is the slope, written here as a function of the expiration tenor $\tau$, but it is also a function of the other model parameters.
	
	We do this to give us a way to estimate how much we expect the risk reversal to move for a given log spot change. If we restrict ourselves to $\rho_0=\bar\rho$ and $v_0=\theta$ for simplicity, we see that
	
	\begin{align*}
	dRR(\tau) &= k(\tau) d\rho_t = k(\tau) \frac{\alpha \eta}{\sqrt{\theta}} dW^\rho_t = k(\tau) \frac{\alpha \eta \rho_{cs}}{\sqrt{\theta}} dW^S_t + ... \\
	 &= \frac{k(\tau) \alpha \eta^2}{\theta} dX_t + ...
	\end{align*}
	
	where the ellipses refer to noise independent of the log return $dX$. Here we used the expression for $\rho_{cs}$ from equation \ref{eq:rho_cs}, and $dW^S_t=dX_t/\sqrt{v_t}$.
	
	In this formulation, then, the risk reversal beta $\beta_{rr}$ is given by the slope,
	
	\begin{equation}
	\beta_{rr} = \frac{k(\tau) \alpha \eta^2}{\theta}
	\end{equation}
	
	This result highlights that this model implies a risk reversal beta that is proportional to $\eta^2$. The term structure of the risk reversal beta is determined by $k(\tau)$, which has no simple closed form, but can be calculated numerically by measuring the slope of change in model-implied risk reversal for small changes in the initial correlation $\rho_0$. It decays on a timescale of about $1/\beta$.

\begin{figure}[htbp]
	\centering
	\includegraphics[width=0.7\textwidth]{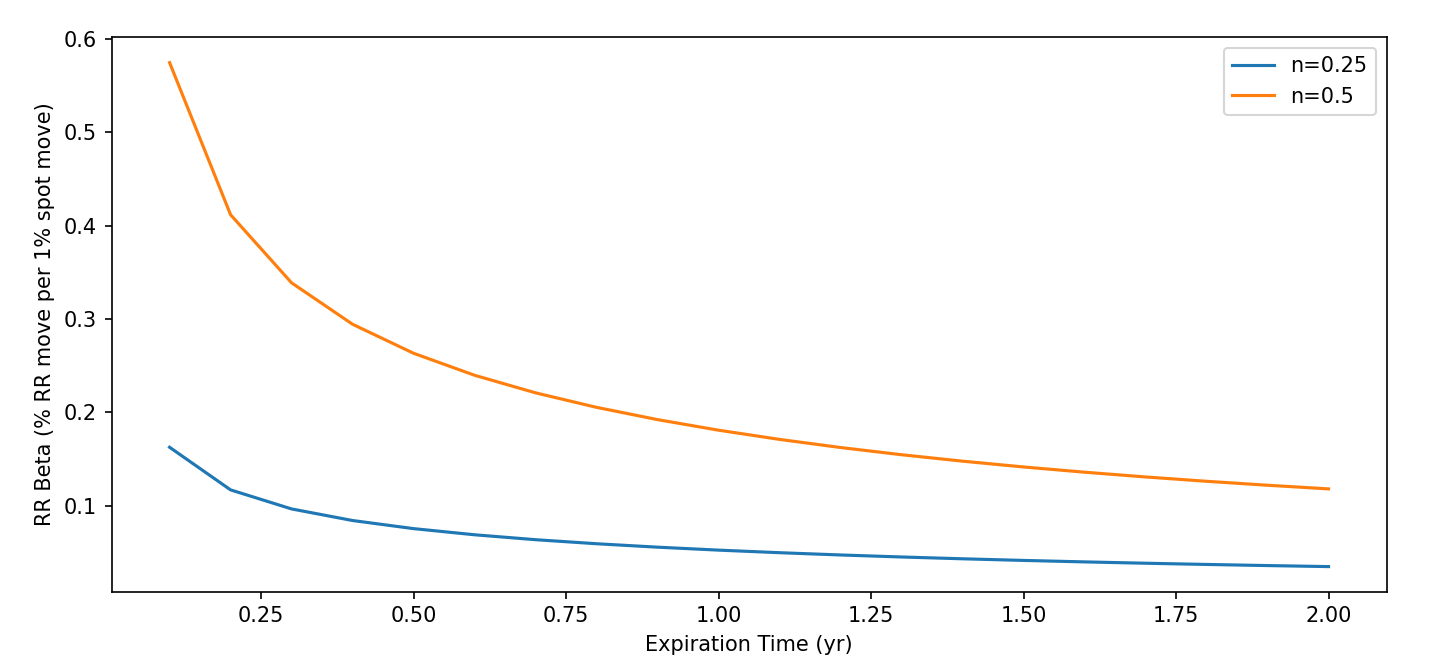}
	\caption{Model Risk Reversal Beta by Option Expiration Tenor}
	\label{fig:rr_beta_by_tenor}
	\vspace{0.5em}
	\begin{minipage}{0.9\textwidth}
		\footnotesize
		The model's risk reversal beta has a term structure that is determined by the value of $\beta$, with a level determined by $\eta^2$. The chart shows the model risk reversal beta for two different values of $\eta$. For this the model was calibrated independently to the same at-the-money volatility, 25-delta risk reversal, and 25-delta butterfly values for all tenors.
	\end{minipage}
    \end{figure}

	Figure \ref{fig:rr_beta_by_tenor} shows a chart of the model's risk reversal beta as a function of expiration tenor, for $\eta=0.25$ and $\eta=0.5$. This was calibrated (tenor by tenor, independently) to a market where the at-the-money volatility is 8.0\%, the 25-delta risk reversal is +1.0\%, and the 25-delta butterfly is 0.5\% - the same for all tenors. We used $\beta=2$.
	
    \subsection{Estimating $\eta$ from Risk Reversal Beta}

    We can now estimate parameter values: if we use an empirical risk reversal beta for \mbox{EURUSD} 3m options of 0.16, use $\theta=0.01$ as a representative variance level, and use $\alpha=0.3$ as a representative volatility of volatility, we can numerically calculate $k(\tau) = 0.037$. Then $\eta = \sqrt{(\theta \beta_{rr}) / (k(\tau) \alpha)} \approx 0.4$ is one estimate of the stochastic correlation parameter for the EURUSD market.
    
    This method of empirically estimating $\eta$ based on the historical dynamics of risk reversals is one way of estimating this parameter's value. Another is to imply it from the market prices of exotic derivatives that depend on it, such as barrier derivatives.
	
	\section{Barrier Option Pricing and the Risk Reversal Beta}
	
	Barrier derivatives are popular products in the foreign exchange markets, often trading as knockout options (vanilla options whose payoffs go to zero if a continuous barrier is breached by the asset spot price any time before expiration) and one touch options (digital options where the owner is paid one unit of currency only if a continuous barrier is breached any time before expiration).
	
	Bid/ask spreads on ``out-of-the-money'' knockout options, where the option is out of the money when the barrier is breached, are comparable to bid/ask spreads on vanilla options in the foreign exchange markets - so very tight indeed, typically measured in a few basis points of the knockout's asset currency notional. Bid/ask spreads on one touch options, whose prices are quoted in percent of the payoff, are typically one or two percent. Accurate pricing models are important when bid/ask spreads are tight on exotic derivatives.

    \subsection{Model Barrier Pricing}

	Unlike with vanilla option pricing, we have found no computational shortcut for barrier option pricing. We use Monte Carlo simulation below to price these derivatives, after calibrating the model to market vanilla option implied volatilities using the fast vanilla option pricing.

    We price continuously monitored barrier claims under our model via Monte Carlo simulation of the log-asset with CIR variance evolved by \cite{Andersen2008HestonQE} quadratic-exponential (QE) scheme (with non-negativity enforced). Barrier crossing is detected each time step using a log-space Brownian-bridge crossing probability that uses expected volatility over the interval. We also use a Feller-ratio-based heuristic that increases the time-step resolution when $\frac{\alpha^2}{2 \beta \theta}$ is large.

    We use the underlying vanilla option as a control variate for out-of-the-money knockout options. For one touches we use a European digital with strike equal to the one touch barrier level as a control variate.
	
	\subsection{One Touch Options}
	
	The risk reversal beta is a key dynamic for pricing one touch options. We can build intuition on this front by considering a hedging strategy for a long one touch position: selling twice the notional of a European digital option with strike equal to the one touch option's barrier level, and sharing the same expiration date. An up-barrier one touch would be hedged by selling twice the notional of a European digital call option, and a down-barrier one touch would be hedged by selling twice the notional of a European digital put option.
	
	Consider an up-barrier one touch option, hedged with twice the notional of a European digital call option. If spot never touches the barrier before the expiration, the one touch ends up worthless - as does the European digital call option. If spot does drift up and touch the barrier before expiration, the one touch pays off and is worth one unit of the payoff currency. The European digital option does not pay off yet - but because spot is at the digital option strike price, assuming small risk neutral drift, the digital has roughly even odds of ending in the money and paying two units of the payoff currency. That means the European digital price is around 50\%, and so the product of twice the notional and a 50\% price gives a value of the European digital that roughly offsets the payoff of the one touch option. Therefore twice the notional of the European digital call option is a fairly close cash flow hedge for the up-barrier one touch, and in practice is a good semi-static vega and gamma hedge as well - semi-static in the sense that the hedge needs to be adjusted - unwound - only if the barrier is breached.
	
	The European digital's price is not exactly 50\% with spot at its strike, however. If we replicate a European digital call option as a tight call spread, its price $P_E$ can be written as
	
	\[
	P_E = -\frac{dC(K)}{dK} = -\frac{\partial C_{BS}(K, \sigma_{iv})}{\partial K} - \frac{\partial C_{BS}}{\partial \sigma_{iv}} \frac{d\sigma_{iv}}{dK}
	\]
	
	Here $C_{BS}(K,\sigma_{iv}(K))$ is the price of a European vanilla call option with strike $K$ and implied volatility $\sigma_{iv}(K)$. 
	
	When spot is at the European digital strike, the first term - the Black-Scholes digital price - is indeed close to 50\%. But the second term shifts the price by an amount proportional to the slope of implied volatility vs strike - that is, the implied volatility skew, or the risk reversal.
	
	If we think back to our portfolio of long one touch, short twice the notional of European digital, when spot touches the barrier, we need to buy back the European digital hedge, since the one touch no longer has any risk to hedge. Per above, the price we buy it back at depends on the risk reversal: the more positive the risk reversal is, the cheaper the European digital is to buy back, so the better it is for us. The total expected hedging cost of the one touch, based on unwinding the European digital hedge conditioned on hitting the barrier at different times, goes down as the risk reversal gets more positive.
	
	This links us back to the risk reversal beta, and why that affects one touch pricing. When we put on the one touch trade, we know the value of the risk reversal; if spot drifts up to the barrier and we expect a positive correlation between moves in risk reversal and moves in spot due to the risk reversal beta, we expect to pay less to buy back the European digital. So the higher the risk reversal beta, the higher the up-barrier one touch price should be.
	
	We can estimate the magnitude of this effect by approximating the unwind cost change due to the risk reversal beta. We approximate
	
	\[
	\frac{\partial C}{\partial \sigma} = k_1 S \sqrt{T}
	\]
	
	where $k_1$ is a constant of order 1, $S$ is the asset spot price, and $T$ is the time to expiration. We approximate the slope of implied volatility vs strike in terms of the risk reversal, approximating the 25-delta strike difference as $k_2 S \sigma \sqrt{T}$:
	
	\[
	\frac{d \sigma}{dK} = k_2 \frac{RR}{S \sigma \sqrt{T}}
	\]
	
	where $k_2$ is another constant of order 1 and $RR$ is the 25-delta risk reversal for expiration $T$.
	
	That means the European digital price correction is 
	
	\[
	- \frac{\partial C_{BS}}{\partial \sigma_{iv}} \frac{d\sigma_{iv}}{dK} = -k_1 k_2 \frac{RR}{\sigma}
	\]
	
	For our purposes we care how much this price correction changes due to the risk reversal beta. Call this extra price correction, on top of what you would expect from a model like Heston where the risk reversal beta is approximately zero, $\Delta P_E$:
	
	\[
	\Delta P_E = k_1 k_2 \frac{\beta_{rr} \ln(B/S)}{\sigma}
	\]
	
	The expected extra unwind cost due to the risk reversal beta is (roughly) this quantity multiplied by the probability of hitting the barrier, which is just the one touch price itself, also of order 1.
	
	Therefore, we can estimate the sign and magnitude of the one touch price correction as approximately $P_{ot} \Delta P_E$.
	
	Consider a market where the risk reversal beta $\beta_{rr} = 0.15$, at-the-money volatility is 8\%, the barrier is 5\% above spot, and the price of the one touch is 50\%. $k_1 \approx 0.4$ and $k_2 \approx 0.7$. The price correction is then around 1.5\%. So we should expect a price impact of the order of a few percent (of the one touch payout amount) from this effect.
	
	One can follow a similar argument for down-barrier one touches and their European digital put option hedges and find the same effect: the price impact of a positive risk reversal beta should be positive, and has the same approximate expression.
	
	\begin{figure}[htbp]
		\centering
		\includegraphics[width=0.7\textwidth]{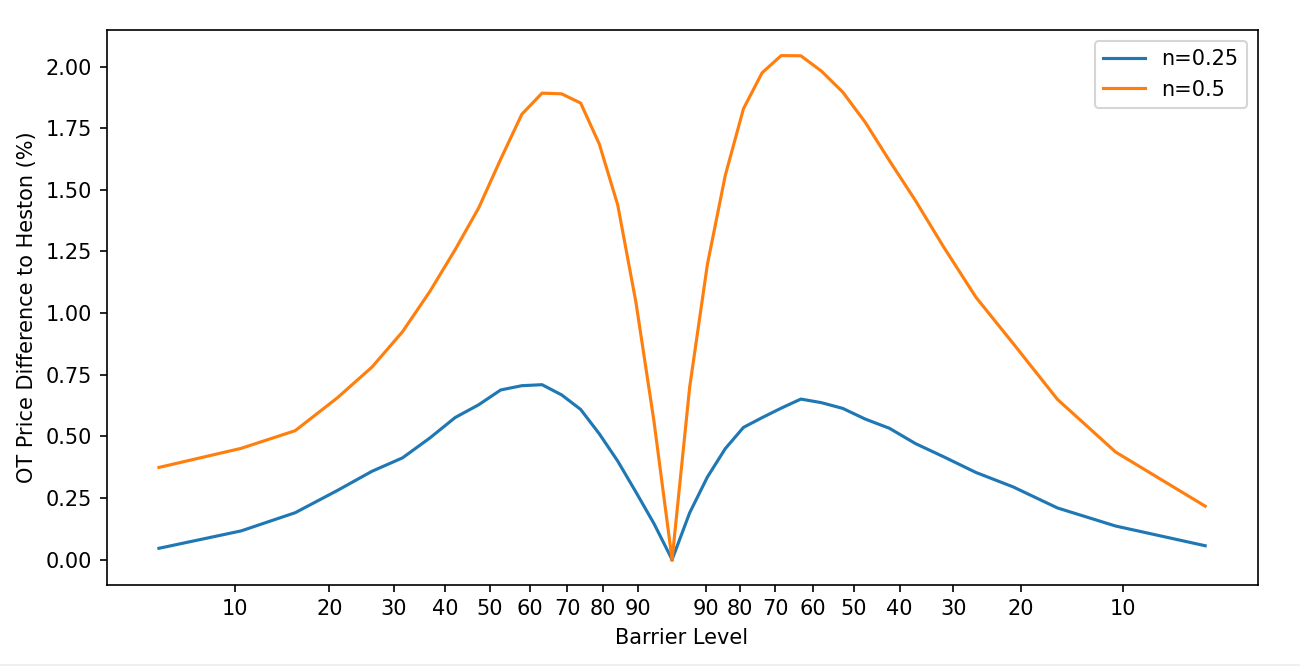}
		\caption{One Touch Price Differences to Heston by Barrier Level}
		\label{fig:ot_by_barrier}
		\vspace{0.5em}
		\begin{minipage}{0.9\textwidth}
			\footnotesize
			This chart shows the difference between our model price for a one touch and its Heston model price, as a function of barrier level. On the x-axis we show, rather than barrier level, the equivalent Black-Scholes one touch price, with the current asset spot price in the center. That is, the barrier level of the one touch is set such that the Black-Scholes one touch price matches the target price. Increasing $\eta$ shows the expected positive price impact to one touch prices, with magnitude near our approximate estimate.
		\end{minipage}
	\end{figure}
	
	Let us compare this rough estimation with the actual results from the model. Figure \ref{fig:ot_by_barrier} shows one touch prices as a function of barrier level. The x-axis shows the barrier strike, but in terms of the Black-Scholes price of one touches - that is, pricing them assuming a constant volatility equal to the at-the-money volatility. The y-axis shows the difference between our model price and the Heston model price, which represents the price impact of stochastic spot/volatility correlation - calibrating model parameters to match the market implied volatilities in both cases. Notice that the model prices always move higher as $\eta$ increases, as expected by the hedging analysis. Also notice that the scale of the impact from stochastic correlation matches our order of magnitude estimate: that is, a few percent.
	
\begin{figure}[htbp]
	\centering
	\includegraphics[width=0.7\textwidth]{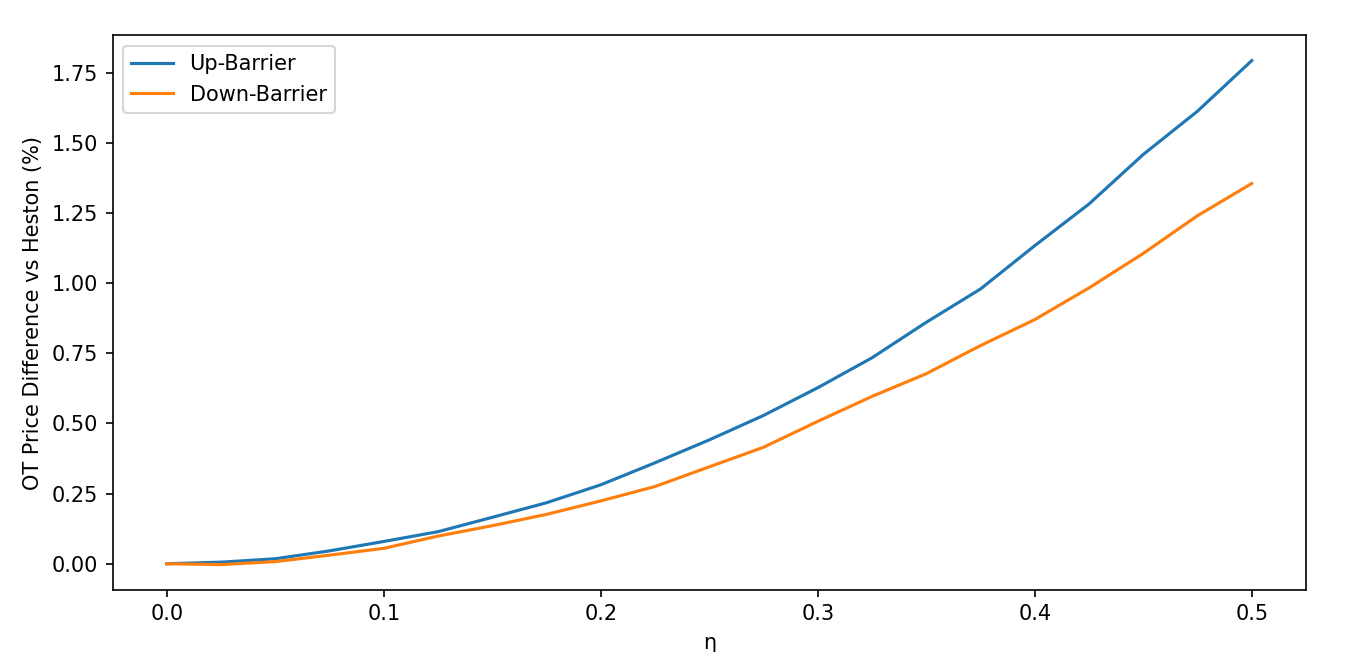}
	\caption{One Touch Price Differences to Heston by \texorpdfstring{$\eta$}{eta}}
	\label{fig:ot_by_eps}
	\vspace{0.5em}
	\begin{minipage}{0.9\textwidth}
		\footnotesize
		This chart shows the difference between the model price for two one touches and their Heston model prices, as a function of $\eta$. The ``Up-Barrier'' line shows the price difference for a 3m-expiration up-barrier one touch with a Black-Scholes price of 50\%, and the ``Down-Barrier'' line shows the price difference for a 3m down-barrier one touch with a Black-Scholes price of 50\%.
	\end{minipage}
	\end{figure}

	The price impact of stochastic correlation is proportional to the risk reversal beta, so should be roughly quadratic in $\eta$. Figure \ref{fig:ot_by_eps} shows this behavior for a 3m-expiration up-barrier one touch with a Black-Scholes price of 50\%, and a 3m-expiration down-barrier one touch with a Black-Scholes price of 50\%, for $\eta$ between 0 and 0.5.
	
	A consequence of this analysis is that using the Heston model to price one touch options, after calibrating the model to vanilla options, leads to one touch prices that are too low, by an amount comparable to or larger than the bid/ask spread. A market maker who uses that model to price new deals and manage risk will find that they tend to accumulate a short position in one touches: their desk tends to make prices that are too low, so counterparties tend to buy from the market maker more often than they sell to them. The trading desk may accumulate quite a large short position until they discover that their model prices are below market prices; when they correct their model they will finally realize their loss as they increase their model prices of positions which they are short.
	
	\subsection{Out-of-the-Money Knockout Options}
	
    A similar hedging argument can be used to demonstrate that the risk reversal beta has a significant positive impact on knockout option prices. 
	
	An ``out-of-the-money'' knockout option is one where the underlying option is out of the money when the barrier is hit. For example, a ``down-and-out'' knockout call option is one where the option knocks down and out - the barrier is below the current spot price - and the barrier level is below the option strike price. Similarly, an ``up-and-out'' knockout put option is one where the option knocks up and out, and the barrier level is above the option strike price.

	Consider a long position in a down-and-out knockout call option, with strike $K$ and barrier $B<K$. A decent semi-static hedge for this derivative is to sell a call option with strike $K$ and buy a put option with strike $K'=B^2/K$, which is less than $B$. If spot never trades down through the barrier, the vanilla call option exactly hedges the expiration cashflows of the knockout call. If, though, spot does trade down to the barrier, the knockout option price goes to zero. At that point, the put option in the hedge is roughly as much out-of-the-money as the call option, so, assuming small risk neutral drift, the call and put prices will be close. The hedge portfolio is short the call and long the put, so if the two prices are similar, the hedge portfolio's price is also close to zero when spot is at the barrier. This two-vanilla portfolio is a decent cashflow hedge for the knockout, and roughly hedges vega and gamma until the barrier is breached.
	
	However, like with the European digital, the price of the two option hedge portfolio is not exactly zero when spot touches the barrier. At that point, as part of this semi-static hedging strategy, we would need to buy the call and sell the put, at whatever the market prices are for the options at that point. 
	
	When spot is at the barrier, both options are roughly the same distance out of the money. That means the vega of the two options also roughly cancels, like the price does: the cost of unwinding that hedge does not depend much on the level of at-the-money volatility when the barrier is breached. Similarly, the risk to the implied volatility smile - the butterfly - roughly cancels off, and the unwind cost does not depend on the butterfly either. However, the portfolio looks like a short risk reversal position for whatever delta the options have when the barrier is breached - which means that the closeout costs do depend sensitively on the level of the risk reversal when the barrier is hit. To close out the hedge we must buy back that risk reversal option position: the more positive the risk reversal, the more we have to pay to unwind the portfolio.
	
	Our best estimate for the level of the risk reversal when spot trades down to the barrier is the current market level less a shift (as spot goes down to the barrier) due to the risk reversal beta. That is, we expect the risk reversal when spot goes to the barrier to be more negative than the current level. The larger the risk reversal beta, the more negative the risk reversal will be when the barrier is breached, and the cheaper it is to buy back the option hedge - which therefore increases the price of the knockout option.

    We can again estimate the magnitude of this effect. This is trickier than for the one touch case, because we do not know the delta of the options when spot touches the barrier. That affects both the vega of the options - and hence their sensitivity to risk reversal moves - and the scale factor for the risk reversal beta. Remember that the risk reversal betas we saw above were for 25-delta options; the risk reversal beta of a risk reversal with delta less than 0.25 is larger than that for a 25-delta option, and the risk reversal beta of a risk reversal with delta greater than 0.25 is smaller.

    Each option's vega is 

	\[
	\mathrm{Vega} = k_1 S \sqrt{\frac{T}{2 \pi}}
	\]

	where $k_1$ depends on the delta of the option, and is between 0 and 1; for a 25-delta option $k_1 \approx 0.4$. We can also write the move in risk reversal as

    \[
    \Delta RR = k_2 \beta_{rr} \ln(S/B) 
    \]

    where $k_2$ depends on the delta of the option, since the risk reversal in question is that for the delta of the options when spot touches the barrier. For deltas less than 0.25, $k_2>1$; for deltas greater than 0.25, $k_2<1$.

    Then the price correction due to the move in risk reversal from the risk reversal beta is

    \[
    \frac{\Delta P_{van}}{S} = k_1 k_2 \beta_{rr} \ln(S/B) \sqrt{\frac{T}{2 \pi}}
    \]

    For a rough estimate, consider the case when the options are 25-delta at the barrier, so $k_1=0.4$ and $k_2=1$. Using a risk reversal beta of 0.16, a barrier 5\% below spot, and 0.25 years left to expiration, the price correction is approximately 6 basis points of the asset currency notional. As with one touches, that correction needs to be multiplied by the probability of hitting the barrier - if that is 50\%, then the price impact is 3 basis points. That is comparable to or larger than the bid/ask spread for out-of-the-money knockout options in the foreign exchange inter-bank market.

    A similar argument can be used to estimate the price impact of the risk reversal beta on up-and-out knockout put options; the price impact is approximately the same, and also positive.

	\begin{figure}[htbp]
	\centering
	\includegraphics[width=0.7\textwidth]{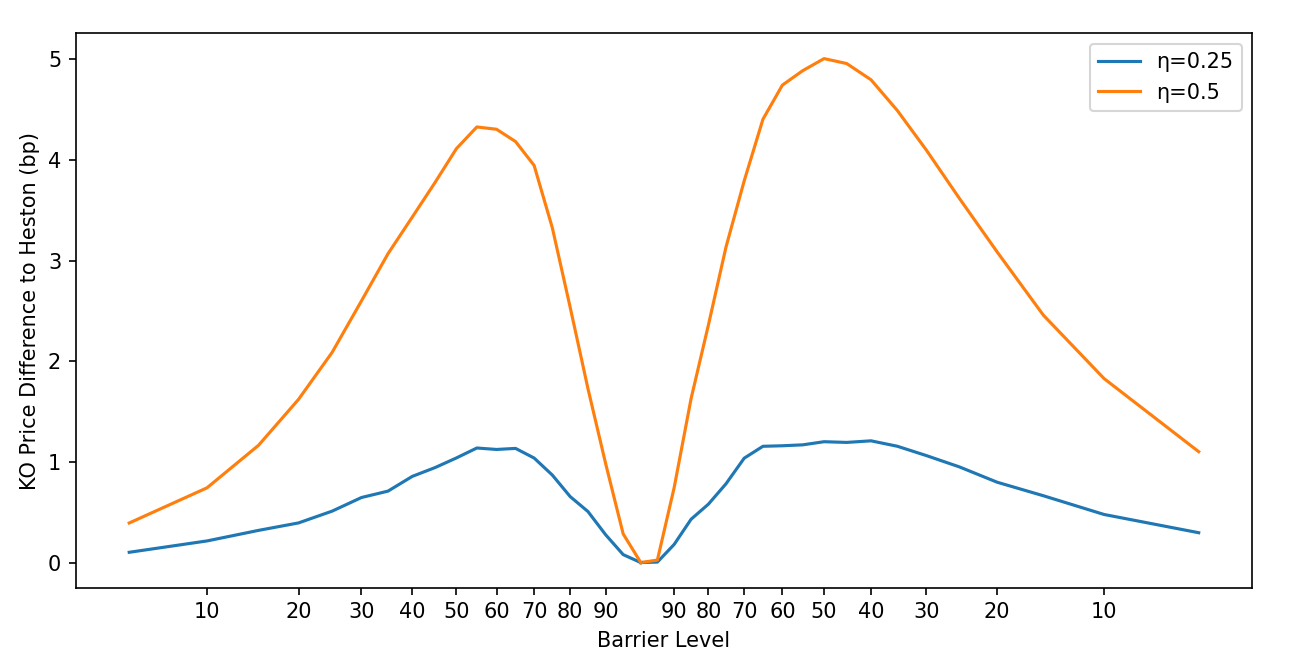}
	\caption{Knockout Price Differences to Heston by Barrier Level}
	\label{fig:ko_by_barrier}
	\vspace{0.5em}
	\begin{minipage}{0.9\textwidth}
		\footnotesize
		This chart shows the difference between the model price for an out-of-the-money knockout option and its Heston model price, as a function of barrier level, in basis points of asset currency notional. On the x-axis we show, rather than barrier level, the equivalent Black-Scholes one touch price, with the current asset spot price in the center. That is, the barrier level of the one touch is set such that the Black-Scholes one touch price matches the target price. Increasing $\eta$ shows the expected positive price impact to knockout prices, with magnitude near our approximate estimate.
	\end{minipage}
	\end{figure}
	
	Figure \ref{fig:ko_by_barrier} quantifies this in our model, showing how knockout prices are affected by $\eta$ at different barrier levels. For barrier levels above spot we use up-and-out knockout put options, and for barrier levels below spot we use down-and-out knockout call options. All options have a strike equal to the initial spot price - that is, they are at-the-money options. The barrier levels are displayed in terms of the Black-Scholes one touch prices as with the one touch chart. The y-axis shows the model price difference to the Heston model price. The price impact is positive as we expected from the hedging analysis, and the price impact is comparable to the approximate estimates we made above.
	
	As with one touches, the price impact of adding stochastic correlation is always positive, if the model is always calibrated to market vanilla option prices. So a market maker using the Heston model to price out-of-the-money knockouts risks finding themselves selling knockout options at what they only discover later to be below-market prices.

    \section{Volatility Swap Fair Strikes and the Risk Reversal Beta}

    \subsection{Volatility Swaps}

    Volatility swaps are another example of liquid exotic derivatives in the foreign exchange markets. Bid/ask spreads are quoted in volatility terms and are typically almost as tight as bid/ask spreads on vanilla option implied volatilities, around 0.1\%. Volatility swaps cannot be statically hedged with vanilla options and so their prices are sensitive to market dynamics not captured in vanilla option prices.
    
    A volatility swap pays the realized volatility of an asset, computed from a set of published once-per-day spot fixings, and exchanges it against a fixed strike set at trade inception. With unit notional, the payoff $P_{vs}$ is

    \[
    P_{vs} = \sigma_r - \sigma_K
    \]

    where $\sigma_K$ is the fixed level of the volatility swap and $\sigma_r$ is the realized volatility:

    \[
    \sigma_r = \sqrt{\frac{N_d}{N} \sum_{i=1}^N R_i^2}
    \]

    Here, $N_d$ is the contract-specified number of trading days per year, which annualizes the daily realized volatility. $N$ is the number of returns in the calculation. $R_i$ is the return between consecutive fixings, computed as the log return of the spot price on day $i$ versus the previous fixing:

    \[
    R_i = \ln \frac{S_i}{S_{i-1}}
    \]

    Fixing times $t_i$ are typically daily. There are $N+1$ spot fixings for the $N$ returns\footnote{This is a simplified realized-volatility definition; some contracts use variations (for example, de-meaning returns before squaring). We will use this simplified form for our analysis.}.

    The ``fair strike'' for a volatility swap is the value of $\sigma_K$ that makes the swap have zero value at inception; equivalently, it is the strike that makes the (discounted) risk-neutral expected payoff $P_{vs}$ equal to zero. The tenor of the volatility swap is the time to the final spot fixing.

    \subsection{Volatility Swaps and Volatility of Volatility Sensitivity}

    A related product to a volatility swap is a variance swap, which swaps that realized volatility \emph{squared} against a fixed level. Variance swaps can be statically\footnote{Approximately hedged, assuming continuous fixings and no jumps in the underlying spot price process.} hedged by a portfolio of vanilla options \citep{DemeterfiDermanKamalZou1999}, which means that the variance swap fair strike should not change when we add stochastic spot/volatility correlation to the model, assuming the model remains calibrated to market vanilla option prices.

    That means we can treat the variance swap as a separate asset, fully specified by the vanilla options market, and consider the volatility swap as a derivative of the variance swap that has a square root payoff. That convex payoff means that volatility swaps cannot be statically hedged by a portfolio of vanilla options, so their prices can depend on market dynamics that are not pinned down by vanilla prices.

    The key market dynamic that affects a volatility swap fair strike is the volatility of volatility. To build intuition, consider a dynamic hedging strategy for a long volatility swap: sell a variance swap with the same fixing schedule, with notional $1/(2\sigma_K)$, to hedge the first-order sensitivity to moves in the variance swap fair strike (which is effectively the underlying asset price for the volatility swap). As volatility rises, the volatility swap's vega is roughly unchanged, while the variance swap's vega increases because of its quadratic payoff. The combined position therefore becomes net short vega, and the trader must buy additional variance swap at the now-higher volatility level to re-hedge. When volatility falls, the variance swap's vega decreases and the trader must sell more variance swap at the lower level. In other words, the hedge is short convexity in the variance swap fair strike - a kind of volatility gamma - so rebalancing losses increase as the volatility of volatility increases.

    \subsection{Volatility Swap Fair Strikes and Stochastic Spot/Volatility Correlation}

    This suggests that the volatility swap fair strike should increase as the volatility of volatility decreases, because these short ``volatility gamma'' hedging costs are decreased. As we saw above, increasing $\eta$ and recalibrating to market vanilla prices means that $\alpha$ - the volatility of volatility parameter in our model - decreases. Therefore the volatility swap fair strike should increase as $\eta$ increases.

	\begin{figure}[htbp]
        \centering
        \includegraphics[width=0.7\textwidth]{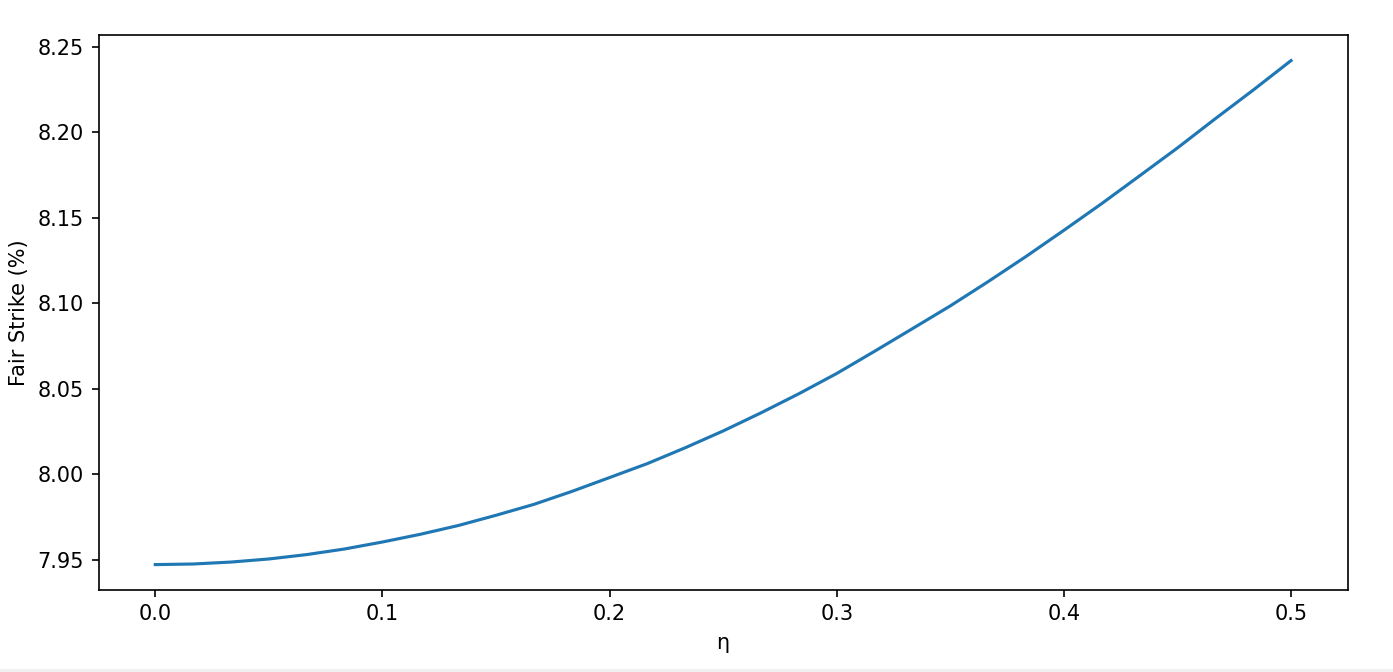}
        \caption{Volatility Swap Fair Strikes vs Volatility of Correlation}
        \label{fig:vs_by_eta}
        \vspace{0.5em}
        \begin{minipage}{0.9\textwidth}
            \footnotesize
            Increasing the volatility of correlation parameter $\eta$ increases the fair strike of a 3m volatility swap because the model needs less volatility of volatility from $\alpha$ to match the market implied volatilities.
        \end{minipage}
        \end{figure}

    Figure \ref{fig:vs_by_eta} shows the volatility swap fair strike as a function of $\eta$. The x-axis shows $\eta$, and the y-axis shows the fair strike for a volatility swap with a 3-month horizon and daily spot fixings. The fair strike increases as $\eta$ increases, consistent with the hedging argument above. We used $N_d=250$ as a representative value. The model is calibrated to an at-the-money volatility of 8.0\%, a 25-delta risk reversal of +1.0\%, and a 25-delta butterfly of +0.5\%.

    Note that a volatility swap price does not explicitly directly depend much on the risk reversal beta dynamic: if we keep other model parameters fixed and only change $\eta$, the volatility swap fair strike hardly changes. The effect displayed in Figure \ref{fig:vs_by_eta} is due to recalibrating the model parameters at different values of $\eta$ to match a fixed implied volatility smile, which changes the calibrated volatility of volatility parameter $\alpha$.

    The size of this pricing adjustment is comparable to or larger than the bid/ask spread for volatility swaps, and it is always positive because increasing $\eta$ always decreases the (calibrated) volatility of volatility $\alpha$. As with out-of-the-money knockout options, market makers who use the Heston model for volatility swap pricing risk selling volatility swaps below market levels.

    For this analysis we calculated volatility swap fair strikes using Monte Carlo simulation, following the same approach as the one touch and knockout option pricing algorithm.

  \section{Related Literature and Model Context}
    \label{sec:related}
    
    Our model is an extension of the Heston stochastic volatility model \citep{Heston1993}, which itself extends the Black-Scholes framework \citep{BlackScholes1973} by introducing a stochastic volatility. Our goal was to add stochastic correlation between spot and volatility, where that correlation is itself correlated with the spot price, and lets us model risk reversal beta.

    There are several other approaches in the literature that incorporate the risk reversal beta dynamic for pricing exotic derivatives, or follow a similar approach to our model.

    \subsection{The Double Heston Model}

    Our model is a version of the Double Heston model defined in \citep{Christoffersen2009}. That model included a second sub-variance not to explicitly introduce stochastic spot/volatility correlation (though it does, and the paper investigates that quantity), but to better match the full implied volatility surface with a small set of constant model parameters. That is, it is more geared at matching implied volatilities across different expiration tenors than focused on the impact of stochastic spot/volatility correlation on exotic derivative prices.
	
	\subsection{Stochastic Correlation in Heston-Type Models}
	
	A natural way to enrich Heston is to relax the assumption of constant spot/volatility correlation and allow correlation itself to be stochastic by explicitly specifying the process it follows. For example, \citet{TengEhrhardtGunther2016} study Heston-type models with stochastic correlation (including Ornstein-Uhlenbeck and bounded ``Jacobi''-type specifications for correlation), and \citet{TengEhrhardtGunther2018} investigate numerical simulation and option pricing implications under stochastic correlation dynamics. These are not affine models and do not admit closed-form characteristic functions, and therefore are more difficult to use in practice: model calibration involves generating many vanilla option prices with different parameters during a numerical rootfinding, and if vanilla option pricing is slow, that calibration process becomes unwieldy. They do admit affine approximations.
	
	Our model differs by making the variance the sum of two sub-variance processes that have different spot/volatility correlations. This allows the spot/volatility correlation to be stochastic, and to be bounded by construction, but still maintains the affine structure of the Heston model.
	
	\subsection{Stochastic Volatility/Local Volatility Mixture Models}
	
	Another common approach to modeling risk reversal beta is to combine stochastic volatility with a deterministic local volatility . These models are often called \emph{local stochastic volatility} (LSV) or \emph{stochastic-local volatility} (SLV) models. A standard specification is
	\[
	\sigma_{\text{inst}}(t,S_t) = \sigma_L(t,S_t)\,\sqrt{v_t},
	\]
	where $v_t$ is a Heston/CIR-type variance factor and $\sigma_L(t,S)$ is the ``local volatility'': a deterministic function of spot $S_t$ and calendar time $t$. In the pure stochastic volatility limit of this model, $\sigma_L$ is a constant and $v_t$ drives the implied volatility skew and smile; in the pure local volatility limit, $v_t$ is deterministic and $\sigma_L$ drives the smile.
	
	An example of a paper that explicitly develops this LSV framework and discusses both vanilla and exotic pricing is \citet{LiptonGalLasis2014}. A closely related and widely used practical incarnation is the ``Heston stochastic-local volatility'' (HSLV) model, in which the Heston dynamics are enhanced by a nonparametric local-volatility component; see \citet{VanDerStoepGrzelakOosterlee2014} for an efficient Monte Carlo approach and discussion of calibration.
	
	These mixture models are regularly used in foreign exchange markets because the local volatility and stochastic volatility limits of the mixture deliver very different risk reversal betas. In a pure stochastic volatility model like the Heston model, the risk reversal is roughly constant since the spot/volatility correlation is constant, and the risk reversal beta is close to zero. In a local volatility model the risk reversal tends to move up and down according to the smile and moves in risk reversal have a high correlation with spot: the risk reversal beta is high. Tuning the mixture parameter lets the model user tune the model's risk reversal beta.
	
	\subsection{Affine Matrix Models and Wishart Stochastic Volatility}
	
	Another common route to stochastic correlation is to model a full stochastic covariance matrix whose implied correlations evolve over time. Wishart (matrix-affine) stochastic volatility models, originating with Wishart processes \citep{Bru1991}, provide affine multivariate dynamics and lead to matrix Riccati equations for transforms; see, e.g., \citet{GourierouxSufana2010}. These approaches work well for modeling multiple correlated assets which all follow Heston processes, but not as well for modeling a single asset with stochastic correlation: we could not find an example of a model that remained affine and also allowed for a separate parameter that controls correlation volatility and risk reversal beta.
	
	\section{Conclusions and Future Work}
	
	We have introduced a variation of the Double Heston model that adds stochastic, mean-reverting spot/volatility correlation to the Heston model. That stochastic correlation is itself correlated with moves in the asset spot price, which we demonstrated is an important dynamic for barrier option pricing.
	
	The model admits fast European vanilla option pricing, aiding in model calibration: like with the Heston model, our model admits a closed-form expression for the characteristic function of the log-spot, and vanilla options can be priced with a single numerical integration.
	
	We derived an expression for the risk reversal beta in this model and compared it to the values we observed in the EURUSD and USDJPY foreign exchange markets. We then used those results to estimate the model parameter $\eta$ that controls the volatility of the spot/volatility correlation.
	
	We also showed how one touch and out-of-the-money knockout option prices in those markets are affected by increasing the volatility of spot/volatility correlation, assuming that the model stays calibrated to a fixed set of vanilla option implied volatilities, and discussed why the sign and magnitude of those effects were expected based on hedging arguments. The price impacts of stochastic spot/volatility correlation are comparable or larger than the bid/ask spread for these derivatives.

    We also considered pricing of volatility swaps, another liquid exotic product in the foreign exchange markets. Volatility swap fair strikes are sensitive to the volatility of volatility, and increasing the spot/volatility correlation volatility $\eta$ reduces the volatility of volatility required to match the market implied volatility smile. This means that the fair strike of a volatility swap should increase as $\eta$ increases.

	An interesting consequence of this model is that adding stochastic spot/volatility correlation always increases the prices of one touch and knockout options, assuming the model is calibrated to market vanilla option prices. This should be concerning to market makers using the Heston model to price and manage risk for these derivatives, as they may find themselves selling these derivatives at below-market prices. Similarly, it increases the fair strike of volatility swaps, so market makers using the Heston model to price volatility swaps risk selling them below market levels.

    One path for future work is to generalize the model to have parameters that are piecewise-constant in calendar time to allow a full term structure calibration to implied volatilities of different strikes and expiration dates. The pricing impact of implied volatility term structure on a barrier option can be significant. In addition, generalizing the model to calibrate cleanly to both 25- and 10-delta implied volatilities, instead of just 25-delta volatilities, would be a useful extension, as 10-delta volatilities are regularly quoted in the FX market.
	
	Another path for future work is to look at traded prices of barrier derivatives in the FX inter-bank market and see whether their prices are better approximated with this model than with other kinds of model used in practice to incorporate this effect, such as stochastic volatility/local volatility mixture models.

	\clearpage
	\appendix
	\section*{Appendixes}
	\addcontentsline{toc}{section}{Appendixes}
	
    \section{Closed-Form Solution of the Model Riccati System}\label{app:ode}

    \subsection{Key Definitions and Notation}

    Define the two limiting correlations

    $$
    \rho_+ := \bar\rho+\eta,\qquad \rho_- := \bar\rho-\eta
    $$

    For each sign $\pm$, define the complex constants (depending on $\xi$, the Fourier argument of the characteristic function)
    $$
    b_\pm(\xi) := \beta - i\xi\,\alpha\,\rho_\pm,
    \qquad
    d_\pm(\xi) := \sqrt{b_\pm(\xi)^2+\alpha^2(\xi^2+i\xi)},
    \qquad
    g_\pm(\xi) := \frac{b_\pm(\xi)-d_\pm(\xi)}{b_\pm(\xi)+d_\pm(\xi)}
    $$
    We take the square-root branch such that $\Re(d_\pm(\xi))\ge 0$ (standard for stability in affine models).

    \subsection{\texorpdfstring{Solving the Riccati Equations for $B_\pm$}{Solving the Riccati Equations for B pm}}

    \subsubsection{\texorpdfstring{Put each $B_\pm$ into canonical Riccati form}{Put each B pm into canonical Riccati form}}

    Fix a sign $\pm$ and write $B(\tau):=B_\pm(\tau;\xi)$, $\rho:=\rho_\pm$. The ODE is
    $$
    B'(\tau)= -\tfrac12(\xi^2+i\xi) + (i\xi\alpha\rho-\beta)B(\tau) + \tfrac12\alpha^2 B(\tau)^2
    $$
    Using $b=\beta-i\xi\alpha\rho$ (so $i\xi\alpha\rho-\beta=-b$), this is
    $$
    B'(\tau)= -\tfrac12(\xi^2+i\xi) - b\,B(\tau) + \tfrac12\alpha^2 B(\tau)^2
    $$
    Equivalently,
    $$
    B'(\tau)=\tfrac12\alpha^2 B(\tau)^2 - b\,B(\tau) -\tfrac12(\xi^2+i\xi)
    $$

    \subsubsection{Factorization by the quadratic roots}

    Consider the quadratic polynomial in $B$:
    $$
    Q(B):=\tfrac12\alpha^2 B^2 - b B -\tfrac12(\xi^2+i\xi)
    $$
    Its discriminant is
    $$
    \Delta=b^2+\alpha^2(\xi^2+i\xi)=d^2,\qquad d:=\sqrt{b^2+\alpha^2(\xi^2+i\xi)}
    $$
    The roots are
    $$
    r_1=\frac{b+d}{\alpha^2},\qquad r_2=\frac{b-d}{\alpha^2}
    $$
    Therefore,
    $$
    Q(B)=\tfrac12\alpha^2(B-r_1)(B-r_2),
    \qquad\text{and hence}\qquad
    B'(\tau)=\tfrac12\alpha^2(B-r_1)(B-r_2).
    $$

    \subsubsection{Separation of variables and explicit solution}

    Separate variables:
    $$
    \frac{dB}{(B-r_1)(B-r_2)}=\tfrac12\alpha^2\,d\tau
    $$
    Using the partial fraction identity
    $$
    \frac{1}{(B-r_1)(B-r_2)}=\frac{1}{r_1-r_2}\left(\frac{1}{B-r_1}-\frac{1}{B-r_2}\right)
    $$
    integrate to obtain
    $$
    \frac{1}{r_1-r_2}\ln\!\left(\frac{B(\tau)-r_1}{B(\tau)-r_2}\right)=\tfrac12\alpha^2\,\tau + C
    $$
    Since $r_1-r_2=\frac{2d}{\alpha^2}$, we have $\tfrac12\alpha^2(r_1-r_2)=d$, hence
    $$
    \ln\!\left(\frac{B(\tau)-r_1}{B(\tau)-r_2}\right)=d\tau + C'
    $$
    Impose $B(0)=0$:
    $$
    \frac{-r_1}{-r_2}=\frac{r_1}{r_2}=e^{C'}
    $$
    Thus
    $$
    \frac{B(\tau)-r_1}{B(\tau)-r_2}=\frac{r_1}{r_2}e^{d\tau}
    $$
    Solving this algebraically for $B(\tau)$ and rewriting in the numerically stable $e^{-d\tau}$ form yields
    $$
    \boxed{
    B(\tau)=r_2\,\frac{1-e^{-d\tau}}{1-\frac{r_2}{r_1}e^{-d\tau}}
    =\frac{b-d}{\alpha^2}\,\frac{1-e^{-d\tau}}{1-g e^{-d\tau}},
    \qquad g=\frac{r_2}{r_1}=\frac{b-d}{b+d}
    }
    $$

    \subsubsection{\texorpdfstring{Apply to $B_+$ and $B_-$}{Apply to B+ and B-}}

    Therefore, for $\pm\in\{+,-\}$,
    $$
    \boxed{
    B_\pm(\tau;\xi)
    =\frac{b_\pm(\xi)-d_\pm(\xi)}{\alpha^2}\,
    \frac{1-e^{-d_\pm(\xi)\tau}}{1-g_\pm(\xi)\,e^{-d_\pm(\xi)\tau}},
    \qquad
    b_\pm(\xi)=\beta-i\xi\alpha\rho_\pm
    }
    $$

    \subsection{\texorpdfstring{ClosedForm Solution for $A(\tau;\xi)$}{Closed-form solution for A(tau;xi)}}

    \subsubsection{\texorpdfstring{Reduce to integrating $B_\pm$}{Reduce to integrating B +-}}

    From the first ODE,
    $$
    A(\tau;\xi)=\int_0^\tau \left(i\xi(r-q)+\beta (\theta_+ B_+(s;\xi)+\theta_- B_-(s;\xi))\right)\,ds
    $$
    so
    $$
    A(\tau;\xi)=i\xi(r-q)\tau + \beta \left(\theta_+ \int_0^\tau B_+(s;\xi)\,ds+\theta_- \int_0^\tau B_-(s;\xi)\,ds\right)
    $$

    Thus it suffices to compute $\int_0^\tau B(s)\,ds$ for the generic closed-form $B(s)=\frac{b-d}{\alpha^2}\frac{1-e^{-ds}}{1-ge^{-ds}}$.

    \subsubsection{\texorpdfstring{Explicit primitive of $B$}{Explicit primitive of B}}

    Let $b,d,g$ correspond to one of the signs $\pm$. Define
    $$
    B(s)=\frac{b-d}{\alpha^2}\frac{1-e^{-ds}}{1-ge^{-ds}}
    $$
    A direct rational substitution (set $y=e^{-ds}$) gives the elementary antiderivative
    $$
    \boxed{
    \int_0^\tau B(s)\,ds
    =\frac{b-d}{\alpha^2}\,\tau
    -\frac{2}{\alpha^2}\ln\!\left(\frac{1-g e^{-d\tau}}{1-g}\right)
    }
    $$
    (One checks the constant by evaluating at $\tau=0$, where the logarithm vanishes and the integral is $0$.)

    \subsubsection{\texorpdfstring{Assemble $A(\tau;\xi)$}{Assemble A(tau;xi)}}

    Applying the above to both components,
    $$
    \boxed{
    \begin{aligned}
    A(\tau;\xi)
    &= i\xi(r-q)\tau
    +\beta \sum_{\pm}
    \theta_\pm
    \left[
    \frac{b_\pm(\xi)-d_\pm(\xi)}{\alpha^2}\,\tau
    -\frac{2}{\alpha^2}\ln\!\left(\frac{1-g_\pm(\xi)\,e^{-d_\pm(\xi)\tau}}{1-g_\pm(\xi)}\right)
    \right]
    \end{aligned}
    }
    $$

    \subsection{Final Closed-Form Characteristic Function}

    With $\tau=T-t$ and $\rho_\pm=\bar\rho\pm\eta$, define
    $$
    b_\pm(\xi)=\beta-i\xi\alpha\rho_\pm,\qquad
    d_\pm(\xi)=\sqrt{b_\pm(\xi)^2+\alpha^2(\xi^2+i\xi)},\qquad
    g_\pm(\xi)=\frac{b_\pm(\xi)-d_\pm(\xi)}{b_\pm(\xi)+d_\pm(\xi)}
    $$
    Then the Riccati solutions are
    $$
    \boxed{
    B_\pm(\tau;\xi)
    =\frac{b_\pm(\xi)-d_\pm(\xi)}{\alpha^2}\,
    \frac{1-e^{-d_\pm(\xi)\tau}}{1-g_\pm(\xi)e^{-d_\pm(\xi)\tau}}
    }
    $$
    and
    $$
    \boxed{
    A(\tau;\xi)
    = i\xi(r-q)\tau
    +\beta\sum_{\pm}
    \theta_\pm
    \left[
    \frac{b_\pm(\xi)-d_\pm(\xi)}{\alpha^2}\,\tau
    -\frac{2}{\alpha^2}\ln\!\left(\frac{1-g_\pm(\xi)e^{-d_\pm(\xi)\tau}}{1-g_\pm(\xi)}\right)
    \right]
    }
    $$
    Therefore,
    $$
    \boxed{
    \mathbb E\!\left[e^{i\xi X_T}\mid\mathcal F_t\right]
    =
    \exp\!\left(
    A(\tau;\xi)+B_+(\tau;\xi)v_t^+ + B_-(\tau;\xi)v_t^- + i\xi X_t
    \right)
    }
    $$
	
	% --- Bibliography ---
	\bibliographystyle{plainnat}
	\bibliography{references} % Your .bib file
	
\end{document}